\documentclass[10pt,a4paper,twocolumn]{article}

\IfFileExists{arxiv-11-16/preamble.tex}{\def\paperroot{arxiv-11-16/}}{\def\paperroot{}}
\newcommand{\paperfile}[1]{\paperroot#1}
\input{\paperfile{preamble.tex}}
\input{\paperfile{metadata.tex}}

\begin{document}

\twocolumn[{
  \begin{minipage}{\textwidth}
  \centering
  {\fontsize{20}{24}\selectfont\bfseries
    \patchcmd{\papertitle}{:}{:\\[2pt]}{}{}
    \papertitle\par}
  \vspace{10pt}
  \begin{minipage}{\textwidth}
    \centering\fontsize{10}{13}\selectfont
    \paperauthors\par
  \end{minipage}
  \par\vspace{6pt}
  \repositorylink
  \par\vspace{6pt}
  \captionsetup{type=figure}
  \includegraphics[width=0.88\textwidth]{\paperroot 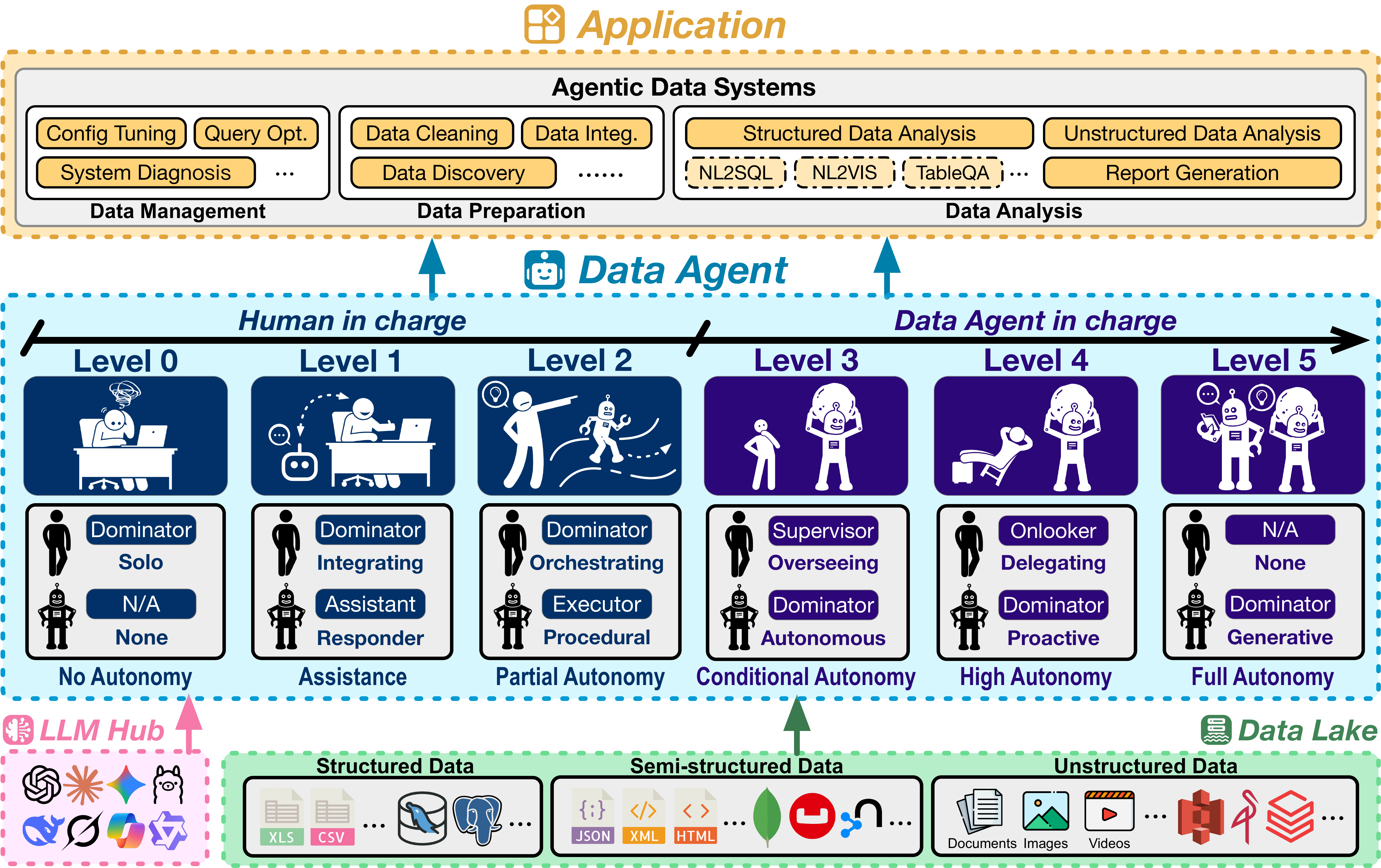}
  \caption{An Overview of Data Agents.}
  \label{fig:overview}
  \vspace{2pt}
  \end{minipage}
}]
\thispagestyle{empty}
\begingroup
\renewcommand{\thefootnote}{}
\makeatletter
\long\def\@makefntext#1{\noindent#1}
\makeatother
\footnotetext{\paperaffiliations}
\endgroup

\begin{abstract}
The rapid advancement of large language models (LLMs) has spurred the emergence of data agents, autonomous systems designed to orchestrate Data + AI ecosystems for tackling complex data-related tasks. However, the term ``data agent'' currently suffers from terminological ambiguity and inconsistent adoption, conflating simple query responders with sophisticated autonomous architectures. This terminological ambiguity fosters mismatched user expectations, accountability challenges, and barriers to industry growth. Inspired by the SAE J3016 standard for driving automation, this survey introduces the first systematic hierarchical taxonomy for data agents, comprising six levels that delineate and trace progressive shifts in autonomy, from manual operations (L0) to a vision of generative, fully autonomous data agents (L5), thereby clarifying capability boundaries and responsibility allocation. 
Through this lens, we offer a structured review of existing research arranged by increasing autonomy, encompassing specialized data agents for data management, preparation, and analysis, alongside emerging efforts toward versatile, comprehensive systems with enhanced autonomy. We further analyze critical evolutionary leaps and technical gaps for advancing data agents, especially the ongoing L2-to-L3 transition, where data agents evolve from procedural execution to autonomous orchestration.
Finally, we conclude with a forward-looking roadmap, envisioning proactive, generative data agents.
A curated companion repository is available at \href{https://github.com/HKUSTDial/awesome-data-agents}{\textit{https://github.com/HKUSTDial/awesome-data-agents}}.
\end{abstract}

\clearpage
\section{Introduction}
\label{sec:intro}

The way humans interact with data is undergoing a revolutionary transformation. Traditionally, handling data-related tasks has been a demanding endeavor, requiring specialized expertise, extensive manual effort, and solid technical proficiency~\cite{datascience_survey, cady2024data, chai2023datamanagement_survey, fernandes2023data, chai2023demystifying, qin2020vis_survey}, fueling a long-standing aspiration in data science and analytics to develop an intelligent agent capable of autonomously managing, preparing, and analyzing data to deliver trustworthy insights with minimal human intervention~\cite{zhou2025survey}.

The advent of large language models (LLMs) and LLM agents is bringing us closer to this vision~\cite{li2024llm, zhou2023llm,li2025data+}. Leveraging their remarkable capabilities in comprehension and reasoning~\cite{minaee2024large, lead, aot}, LLM agents have evolved beyond simple question answering~\cite{liu2025advances}. Recent advances in LLM-based agents demonstrate not only enhanced reasoning ability but also emerging capacities for environmental perception and interaction~\cite{yao2023react, yao2024language}, memory retention~\cite{sumers2023cognitive}, strategic planning~\cite{zhang2024aflow}, and external tool invocation~\cite{zhuang2023toolqa, feng2025retool}. 

\vspace{-0em}
\subsection{The Dawn of Data Agents}

Within this trend, \textit{Data Agents} are introduced to address the distinctive challenges of data-intensive environments. A data agent is defined as a comprehensive, LLM-powered architecture that orchestrates the Data + AI ecosystem to autonomously perform a wide range of data-related tasks~\cite{sun2025data, fu2025autonomous}. 
As illustrated in Figure~\ref{fig:overview}, the data agents serve as a central intelligence layer that bridges user-facing applications with the underlying data infrastructure by producing output such as optimized database configurations, prepared data, data insights, visualization charts, or analytical reports.
Formally, we can define a data agent $\mathcal{A}$ that operates on raw data $\mathcal{D}$ within an environment $\mathcal{E}$ (\eg DBMS, code interpreters, APIs, etc.), utilizing LLMs $\mathcal{M}$, ultimately producing an output $\mathcal{O}$ to tackle the data-related task $\mathcal{T}$, abstractly represented as:
\vspace{-.0em}
\begin{equation*}
\mathcal{A}: (\mathcal{T}, \mathcal{D}, \mathcal{E}, \mathcal{M}) \rightarrow \mathcal{O}.
\end{equation*}
\vspace{-0em}

Different from general-purpose LLM agents, data agents are specialized for data-intensive environments, where data infrastructure serves as the environment for perception, action, and validation rather than merely part of the task context.
As summarized in Table~\ref{tab:comparison_general_data_agents}, data agents interact with DBMSs, data lakes, BI systems, ETL tools, and runtime engines, producing verifiable data artifacts such as SQL queries, database configurations, cleaned or integrated data, visualizations, reports, workflows, and execution traces.

Therefore, data agents must ground their reasoning in the semantics, states, and feedback of data systems, including schemas, values, metadata, constraints, provenance, and execution results. 
Such grounding is essential because errors in data-related tasks are rarely isolated, which can propagate from schema misinterpretation to incorrect queries, flawed preparation, misleading analysis, and downstream decision errors. 
Consequently, data agents require closed-loop capabilities for data exploration, execution-aware validation, repair, provenance tracing, and control of cascading errors.

Addressing these challenges demands specialized capabilities, including: 
(i) interactive perception and exploration of data lakes via environmental feedback; 
(ii) robust invocation of specialized data toolkit, such as SQL equivalence checker, DBMS utilities, code interpreters, or visualization libraries; 
(iii) adaptive resolution and specialized knowledge to dynamically handle noise, inconsistencies, scalability constraints, and real-time updates, preventing errors from cascading.

The emergence of data agents marks a critical step toward realizing the aspiration of democratizing data-related tasks~\cite{li2024llm, tang2025llm, wang2025large}. This progress is evident across the data lifecycle which outlines three interconnected phases: (1) Data Management, encompassing (i) Configuration Tuning for optimizing system parameters (\eg database knobs) to boost performance; (ii) Query Optimization involving SQL rewriting and efficient execution plan selection; and (iii) System Diagnosis for detecting, analyzing, and resolving anomalies or faults. (2) Data Preparation, involving (i) Data Cleaning to detect and fix errors, inconsistencies, or missing values in raw data; (ii) Data Integration to merge heterogeneous sources; and (iii) Data Discovery to identify relevant datasets, metadata, or patterns. (3) Data Analysis, covering (i) Structured Data Analysis for reasoning over tabular/relational data (\eg via TableQA, NL2SQL, or NL2VIS); (ii) Unstructured Data Analysis to extract insights from documents, images, etc.; and (iii) Report Generation to compile findings into coherent narratives.

Recent studies seek to alleviate the expertise and labor-intensive nature of these tasks with advanced data agents. 
Systems like GaussMaster~\cite{zhou2025gaussmaster} and AutoPrep~\cite{fan2025autoprep} are being developed to assist database maintenance and data preparation.
Alpha-SQL~\cite{li2025alphasql} and nvAgent~\cite{nvAgent2025ACL} facilitate seamless interaction with databases and the curation of visualizations through natural language interfaces.
More comprehensive frameworks like iDataLake~\cite{wang2025idatalake} even start to support versatile data operations across heterogeneous data lakes.

\begingroup
\renewcommand{\arraystretch}{1.00}
\begin{table*}[t]
\caption{Comparison between General-purpose LLM Agents and Data Agents}
\label{tab:comparison_general_data_agents}
\centering
\resizebox{0.92\textwidth}{!}{\begin{tabular}{|c|l|l|}
\hline
\textbf{Aspect} &
  \multicolumn{1}{c|}{\textbf{General-purpose LLM Agents}} &
  \multicolumn{1}{c|}{\textbf{Data Agents}} \\ \hline

\begin{tabular}[c]{@{}c@{}}Core Goal\end{tabular} &
  \begin{tabular}[c]{@{}l@{}}
  \textbf{Broad User Tasks:} \\
  Assist with writing, search, planning, dialogue, \\
  and content generation.
  \end{tabular} &
  \begin{tabular}[c]{@{}l@{}}
  \textbf{Data-centric Tasks:} \\
  Support data management, preparation, analysis, operation, \\
  and governance.
  \end{tabular} \\ \hline

\begin{tabular}[c]{@{}c@{}}Operating\\Environment\end{tabular} &
  \begin{tabular}[c]{@{}l@{}}
  \textbf{Open-ended Digital Environments:} \\
  Interact with task-dependent contexts, tools, \\
  webpages, or APIs.
  \end{tabular} &
  \begin{tabular}[c]{@{}l@{}}
  \textbf{Data Infrastructure:} \\
  Operate over DBMSs, data lakes, BI systems, ETL tools, \\
and runtimes.
  \end{tabular} \\ \hline

\begin{tabular}[c]{@{}c@{}}Grounding and\\Interaction\end{tabular} &
  \begin{tabular}[c]{@{}l@{}}
  \textbf{Prompt- and Tool-grounded:} \\
  Reason over prompts, retrieved content, \\
  and tool observations when available.
  \end{tabular} &
  \begin{tabular}[c]{@{}l@{}}
  \textbf{Data-grounded and Execution-aware:} \\
  Ground reasoning in schemas, values, metadata, constraints, \\
provenance, and execution feedback.
  \end{tabular} \\ \hline

\begin{tabular}[c]{@{}c@{}}Tool and\\Execution\end{tabular} &
  \begin{tabular}[c]{@{}l@{}}
  \textbf{General-purpose Tool Use:} \\
  Invoke tools such as search engines, calculators, \\
  OCR, code, or web APIs.
  \end{tabular} &
  \begin{tabular}[c]{@{}l@{}}
  \textbf{Specialized Data Tool Use:} \\
  Invoke SQL engines, DB loaders, profilers, optimizers, \\
  validators, visualization libraries, or code/runtime engines.
  \end{tabular} \\ \hline

\begin{tabular}[c]{@{}c@{}}Outputs and\\Evidence\end{tabular} &
  \begin{tabular}[c]{@{}l@{}}
  \textbf{User-facing Artifacts:} \\
  Generate text, plans, content, answers, \\
  or tool-mediated actions.
  \end{tabular} &
  \begin{tabular}[c]{@{}l@{}}
  \textbf{Verifiable Data Artifacts:} \\
  Produce SQL, configurations, cleaned/integrated data, \\
  charts, reports, workflows, or execution traces.
  \end{tabular} \\ \hline

\begin{tabular}[c]{@{}c@{}}Reliability and\\Evaluation\end{tabular} &
  \begin{tabular}[c]{@{}l@{}}
  \textbf{Direct Task Reliability:} \\
  Emphasize response correctness, task success, \\
  and tool-use robustness.
  \end{tabular} &
  \begin{tabular}[c]{@{}l@{}}
  \textbf{Downstream Workflow Reliability:} \\
  Require data grounding, workflow robustness, auditability, \\
  and control of cascading errors.
  \end{tabular} \\ \hline

\end{tabular}}
\vspace{-0.5em}
\end{table*}
\endgroup

\vspace{-0em}
\subsection{The Terminological Ambiguity of Data Agents}
\label{sec:crisis_of_definition}

Despite significant advancements and promise, the term ``Data Agent'' is applied inconsistently across research and industry, resulting in considerable terminological ambiguity. This ambiguity conflates systems of profoundly different autonomy, reliability, and complexity under a single, imprecisely defined umbrella term. 
For instance, some studies focus on developing sophisticated agentic data systems to autonomously interact with data lakes, orchestrate and optimize tailored pipelines for highly comprehensive and complex data-related tasks with minimal human intervention. In contrast, the ``data agent'' term is also applied to more rudimentary, narrowly scoped systems, often limited to providing atomic task assistance on a ``prompt-response'' basis.

Such terminological ambiguity introduces challenges for user trust, governance, and the healthy progression of this field.

\bi
    \item {\em User-Side Risk.}
        The ambiguity creates expectation mismatches. When users are unable to be clearly aware of the scope and limitations of data agents' capabilities, they can form a flawed understanding of their functionality, leading to a gap between users' expectations and the system's actual performance. Consequently, users may either reject valid outputs or place undue reliance on erroneous ones.
    \item {\em Governance Risk.}
        It further exposes accountability challenges. If a data agent is operated beyond its capabilities and results in negative consequences (\eg data leakage, misleading analysis, regulatory non-compliance), the responsibility becomes unclear. Whether the human operator or system vendor is liable creates complex legal and ethical issues, complicating effective governance.
    \item {\em Industry-Side Risk.}
        User misunderstanding and ambiguous accountability create barriers to industry development. Lacking a shared taxonomy hinders objective comparison and fosters blurring even overstated claims, exacerbating the aforementioned risks. Unclear accountability for failures can erode market confidence and slow technology adoption.
        Addressing these challenges requires establishing a clear, common language for classifying data agents.
\ei

\vspace{-0em}
\subsection{A Hierarchical Taxonomy for Data Agents}
\label{sec:intro_taxonomy}

\begin{figure*}[t!]
  \centering
  \includegraphics[width=0.99\linewidth]{\paperroot 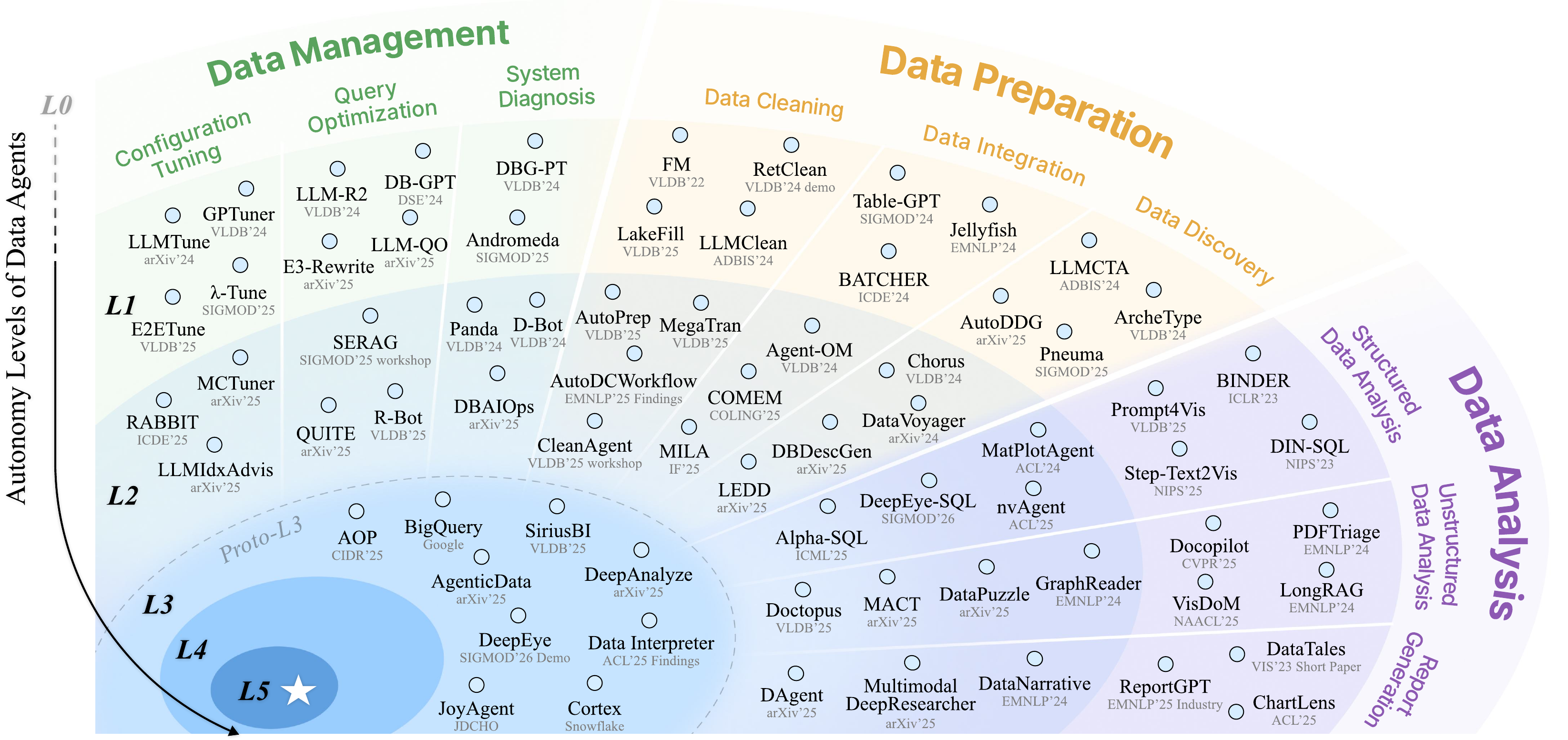}
  \vspace{-.0em}
  \caption{Representative Data Agents Across Different Levels.}
  \label{fig:representative_work}
  \vspace{-0em}
\end{figure*}

The challenges currently facing data agents are not unprecedented. The driving-automation community had faced a similar issue with the term ``self-driving'', which conflated capabilities ranging from driver assistance to high autonomy.
To resolve this, the Society of Automotive Engineers (SAE) introduced the J3016 standard~\cite{SAE-J3016}, a six-level taxonomy. By explicitly allocating responsibility between human and machine, SAE J3016 has established a common language for users, manufacturers, engineers, regulators, and insurers to assess capabilities, limitations, and liability~\cite{ramos2023automated}, which has proven effective in grounding public discourse, clarifying accountability, and guiding engineering roadmaps.

The precedent offers an illuminative path forward. Inspired by this, we propose a hierarchical taxonomy for data agents, analogous to J3016. This framework maps the progressive transition of dominance and responsibility in data-related tasks from humans to data agents as autonomy increases from L0 to L5, reflecting how humans shift from hands-on operators to supervisors and eventually disengagement, while data agents evolve from auxiliary responders toward the vision of fully autonomous and accountable data scientists, as illustrated in Figure~\ref{fig:overview}. A detailed elaboration of each level is provided in Section~\ref{sec:levels}.

Through this taxonomy, we examine existing data agents to structure this survey (Figure~\ref{fig:representative_work}), aiming to clarify capability boundaries and accountability at each level, helping manage expectations, support governance, and steer future research. 

\vspace{-0em}
\subsection{Comparison and Our Contributions}

\vspace{-.0em}

\stitle{Differences from Existing Surveys.}
Existing surveys and tutorials have examined LLMs and agents for data from complementary perspectives, including LLM-data interactions~\cite{zhou2025survey,li2024llm,li2025data+}, data-analyst workflows~\cite{wang2025large,tang2025llm,maojun2025survey,rahman2025llm,wang2025llm4ds}, and task-specific domains such as data management, preparation, and database interfaces~\cite{tan2024large,freire2025large,qin2020vis_survey,chai2023datamanagement_survey,chai2023demystifying,zhou2023llm,liu2025survey,luo2025natural}. 
In contrast, our survey asks a different question: given a data-related task, who orchestrates the workflow, who interacts with the data environment, and who assumes accountability for the task?
We introduce an autonomy-centered L0--L5 taxonomy that maps progressive shifts in orchestration, execution, supervision, and accountability. 
This lens enables a lifecycle-wide review of cutting-edge data agents that remain underrepresented in existing surveys, distinguishes systems with similar components but different capability boundaries, and exposes autonomy gaps and bottlenecks, especially in the L2-to-L3 transition.
Finally, we present a forward-looking roadmap envisioning the path toward proactive and generative data agents.

\vspace{-.0em}
\stitle{Contributions.}
We make the following contributions.
\vspace{-.0em}

\bi

    \item \textit{Novel Hierarchical Taxonomy:}
    Establishing a systematic hierarchical taxonomy for data agents to compare existing works, delineate capability boundaries, and clarify accountability, enabling practitioners to align expectations and intervention with autonomy levels.
        
    \item \textit{Structured and Systematic Review:}
    A structured review of data agents tracing autonomy progression in data-related tasks. Through the lens of our hierarchical taxonomy framework, we map the existing systems, allowing a clear comparison.

    \item \textit{Analysis of Evolutionary Leaps and Gaps:}
    In-depth analysis of evolutionary leaps and current challenges hindering the development of data agents towards higher autonomy levels.

    \item \textit{Forward-Looking Roadmap and Vision:}
    Detailing promising future directions and visions toward trustworthy, proactive, and generative data agents, highlighting opportunities in autonomous orchestration, data lifecycle-wide generalization, reliability, and accountability.
\ei

\vspace{-0em}
\section{Levels of Data Agents} \label{sec:levels}
\vspace{-.0em}

\subsection{The Overview of Hierarchical Taxonomy}
\vspace{-.0em}

To address the terminological ambiguity surrounding ``data agent'', as discussed in Section~\ref{sec:crisis_of_definition} and \ref{sec:intro_taxonomy}, we propose a hierarchical taxonomy from L0 to L5 inspired by SAE J3016~\cite{SAE-J3016}, categorizing data agents based on degree of autonomy.

As shown in Figure~\ref{fig:overview}, the introduced taxonomy is structured around the progressive transfer of dominance and responsibility in data-related tasks from the human to the data agent. When the data agent's autonomy level increases, the human's role transfers from a hands-on dominator to a supervisor, ultimately to disengagement, while the data agent evolves from an auxiliary query responder to a fully autonomous, generative data scientist. 
Below, we define and elaborate on each level.

\bi
    \item \textbf{L0: No Autonomy.}
    All data management, preparation, and analysis tasks are entirely human-driven.
    Humans are solo practitioners, while data agents are uninvolved yet. 
    \item \textbf{L1: Assistance.}
    Emerging from the initial wave of LLMs, 
    L1 data agents act primarily as nascent, stateless query-responsive assistants, offering basic support for data-related tasks by responding to users' queries, while humans retain task dominance and responsibility for interacting with the environment, integrating, and verifying data agents' outputs.

    \item \textbf{L2: Partial Autonomy.}
    At L2, data agents gain the ability to perceive and interact with their environment (\eg data lakes, code interpreters, APIs, etc.). 
    L2 data agents evolve into procedural executors, utilizing memory, external tools, and environmental feedback to adaptively optimize their actions, enabling partial autonomy in task-specific procedures. However, they operate within human-orchestrated pipelines, where humans remain responsible for managing the overall pipeline and still retain dominance over data-related tasks.
    
    \item \textbf{L3: Conditional Autonomy.}
    L3 data agents are expected to autonomously orchestrate tailored data pipelines for a wide range of diverse and comprehensive data-related tasks under supervision, extending beyond human-defined workflows or specific tasks. This marks a critical transition where data agents assume dominance in data-related tasks, while humans act as supervisors overseeing the process.

    \item \textbf{L4: High Autonomy.}
    Human supervision is eliminated. Data agents proactively identify issues worthy of investigation through continuous monitoring and exploration of data lakes, and selectively and reliably orchestrate pipelines to tackle self-discovered problems. Humans delegate full responsibility and become onlookers.

    \item \textbf{L5: Full Autonomy.}
    The ultimate level where data agents go beyond applying existing methods, inventing novel solutions and pioneering new paradigms to advance the state-of-the-art in data management, preparation, and analysis, making any form of human involvement unnecessary.
\ei

\noindent We organize the literature review according to the increasing levels of data agent autonomy in the following sections. More detailed explanations, examples and illustrations are provided in the corresponding sections (Sections~\ref{sec:l0_l1}, \ref{sec:l2}, \ref{sec:l3}, and~\ref{sec:l4_l5}).

\vspace{-0em}
\subsection{The Evolutionary Leaps Between Levels}
\label{sec:leaps}
\vspace{-.0em}

\begin{figure}[t!]
  \centering
  \includegraphics[width=1.0\linewidth]{\paperroot 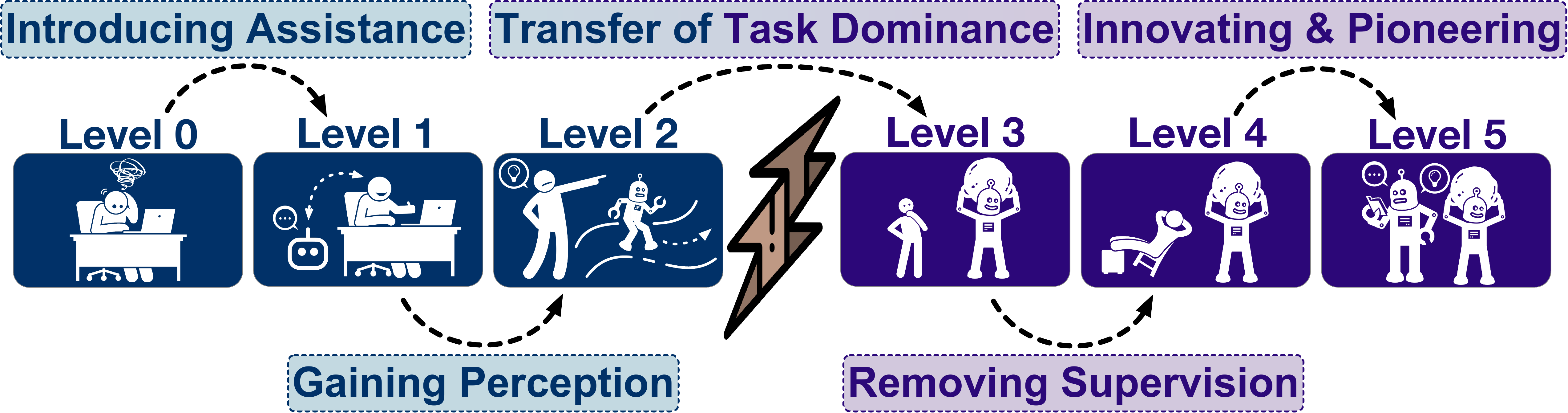}
  \vspace{-0em}
  \caption{ Evolutionary Leaps Between Data Agent Levels.}
  \label{fig:leaps}
  \vspace{-0em}
\end{figure}

The leaps across levels mark data agents' progression in intelligence and autonomy, alongside shifting human–agent roles. Understanding these transitions is key to recognizing the technical and conceptual challenges.

As summarized in Figure~\ref{fig:leaps}, the initial L0$\rightarrow$L1 leap introduces assisted intelligence, shifting humans from solo practitioners to users of query-responsive assistants. The L1$\rightarrow$L2 leap gains data agents with perception, evolving from stateless responders to procedural executors that can perceive data environments and adapt to feedback. Following this, the L2$\rightarrow$L3 leap marks the critical transfer of task dominance, moving data agents from executing human-defined pipelines to autonomous orchestration for versatile and comprehensive data-related tasks. Next, the L3$\rightarrow$L4 leap eliminates supervision, shifting humans to onlookers. The final L4$\rightarrow$L5 leap envisions generative data agents inventing methodologies beyond merely applying existing ones to advance data science and analytics. 

This taxonomy offers a unified lens for analyzing the current landscape of data agents and identifying critical challenges. Building on this foundation, we next review the literature, organized by this hierarchical taxonomy.

\section{L0--L1: From Manual Labor to Preliminary Assistance} 
\label{sec:l0_l1}

In this section, we first review conventional manual processes involved in data-related tasks (L0), followed by the initial advancement represented by the introduction of L1 data agents as preliminary assistance.

\vspace{-0em}
\subsection{L0: Manual Labor in Early Ages}
Conventionally, all data management, preparation, and analysis tasks are performed entirely by humans without intelligent assistance. 
Formally, the human $\mathcal{H}$ is responsible for the entire process, orchestrating ($\pi_{\mathcal{H}}$) pipeline $P$ and executing ($\epsilon_{\mathcal{H}}$), while the data agent $\mathcal{A}$ is uninvolved yet:
\vspace{-.0em}
\begin{equation*}
\begin{aligned}
\mathcal{H} &: \pi_{\mathcal{H}}(\mathcal{T}, \mathcal{D}, \mathcal{E}) \rightarrow P; \quad
\epsilon_{\mathcal{H}}(P, \mathcal{D}, \mathcal{E}) \rightarrow \mathcal{O} \\
\mathcal{A} &: \emptyset
\end{aligned}
\end{equation*}
\vspace{-.0em}

For instance, in data management, humans need to closely monitor databases and manually tune configurable knobs in DBMS (\eg shared buffers and maximum connections), diagnose and resolve issues based on their expertise~\cite{duan2009tuning, zhao2023automatic}. Similarly, data preparation typically involves writing scripts for data cleaning, addressing numerous minor issues during data transformation, and integrating heterogeneous data to support downstream analysis~\cite{fernandes2023data}. During data analysis, analysts often spend considerable time and effort writing SQL queries to extract and analyze relevant data~\cite{tanimura2021sql}, curating visualizations via Vega-Lite or other tools, and meticulously producing insightful interpretations~\cite{satyanarayan2016vega, grammel2010information}.

Such a traditional mode is labor-intensive and time-consuming, requiring extensive domain knowledge and strong technical skills, which creates high barriers for non-experts. Efforts have consistently been made to alleviate these challenges, and this field has changed significantly with the advent of LLMs with powerful reasoning and comprehension abilities, starting to provide preliminary assistance in data-related tasks.

\vspace{-0em}

\subsection{L1: Preliminary Assistance}
\vspace{-.0em}

\begin{figure}[t!]
    \centering
    \includegraphics[width=0.95\linewidth]{\paperroot 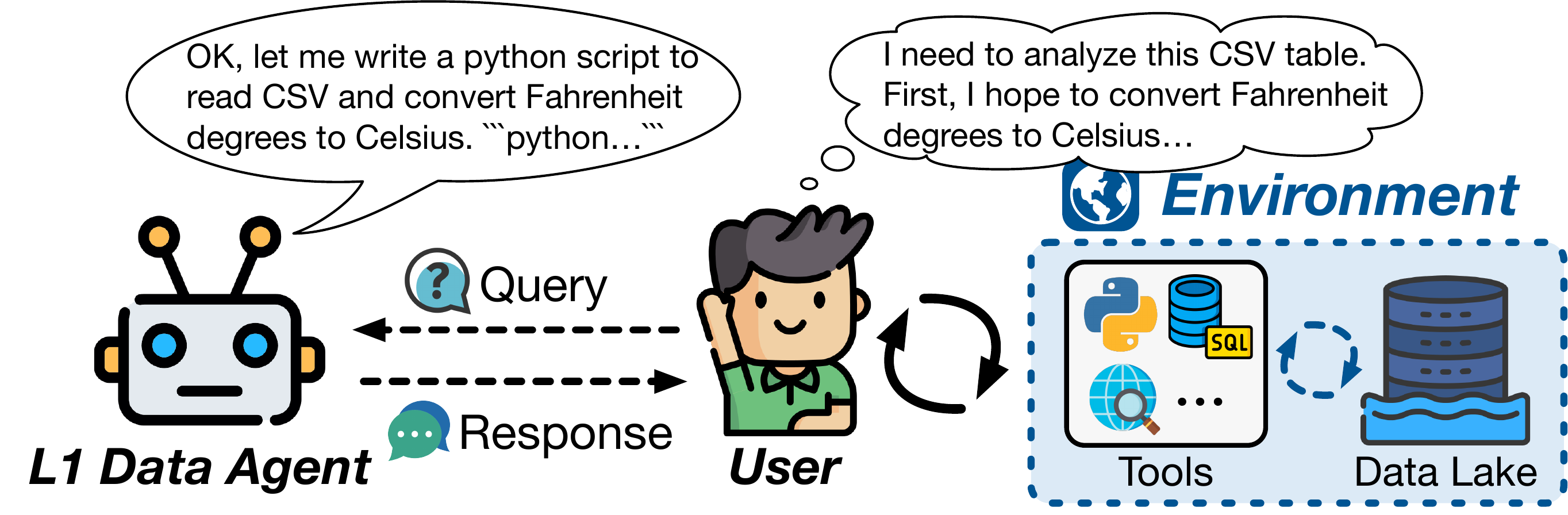}
    \vspace{-.0em}
    \caption{ L1 Data Agents (Assistance).}
    \label{fig:l1_agent}
    \vspace{-0.0em}
\end{figure}

\begin{table*}[t!]
\centering
\caption{Comparison of Representative L1 Data Agents. RAG: Retrieval-Augmented Generation; SFT: Supervised Fine-Tuning; RL: Reinforcement Learning. Data complexity: Multi-source (Multis.), Heterogeneous (Hete.), and Multimodal (Multim.).}
\vspace{-0em}
\label{tab:l1_comparison}
\resizebox{0.87\textwidth}{!}{\begin{tabular}{|c|c|c|c|c|c|c|ccc|}
\hline
\multirow{2}{*}{\textbf{}} &
  \multirow{2}{*}{\textbf{Task}} &
  \multirow{2}{*}{\textbf{Data Agent}} &
  \multirow{2}{*}{\textbf{In-Context Learning}} &
  \multirow{2}{*}{\textbf{RAG}} &
  \multirow{2}{*}{\textbf{SFT}} &
  \multirow{2}{*}{\textbf{RL}} &
  \multicolumn{3}{c|}{\textbf{Data Complexity}} \\ \cline{8-10} 
 &
   &
   &
   &
   &
   &
   &
  \multicolumn{1}{c|}{\textbf{Multis.}} &
  \multicolumn{1}{c|}{\textbf{Hete.}} &
  \textbf{Multim.} \\ \hline
\multirow{10}{*}{\rotatebox{90}{\color[HTML]{000000}Data Management}} &
  \multirow{4}{*}{\begin{tabular}[c]{@{}c@{}}Configuration \\ Tuning\end{tabular}} &
  LLMTune~\cite{huang2024llmtune} &
  zero-shot &
  - &
  \greencheck &
  - &
  \multicolumn{1}{c|}{-} &
  \multicolumn{1}{c|}{\greencheck} &
  - \\ \cline{3-10}  
 &
   &
  GPTuner~\cite{lao2024gptuner} &
  zero-shot &
  - &
  - &
  - &
  \multicolumn{1}{c|}{\greencheck} &
  \multicolumn{1}{c|}{\greencheck} &
  - \\ \cline{3-10}
 &
   &
  $\lambda$-Tune~\cite{giannakouris2025lambda} &
  zero-shot &
  \greencheck &
  - &
  - &
  \multicolumn{1}{c|}{-} &
  \multicolumn{1}{c|}{-} &
  - \\ \cline{3-10} 
 &
   &
  E2ETune~\cite{huang2025e2etune} &
  few-shot &
  - &
  \greencheck &
  - &
  \multicolumn{1}{c|}{-} &
  \multicolumn{1}{c|}{-} &
  - \\ \cline{2-10} 
 &
  \multirow{4}{*}{\begin{tabular}[c]{@{}c@{}}Query \\ Optimization\end{tabular}} &
  DB-GPT~\cite{zhou2024db_gpt} &
  few-shot &
  \greencheck &
  \greencheck &
  - &
  \multicolumn{1}{c|}{\greencheck} &
  \multicolumn{1}{c|}{\greencheck} &
  - \\ \cline{3-10} 
 &
   &
  LLM-$\text{R}^2$~\cite{li2024llm_r2} &
  few-shot &
  \greencheck &
  - &
  - &
  \multicolumn{1}{c|}{-} &
  \multicolumn{1}{c|}{\greencheck} &
  - \\ \cline{3-10}  
 &
   &
  LLM-QO~\cite{tan2025llmqo} &
  few-shot &
  - &
  \greencheck &
  - &
  \multicolumn{1}{c|}{-} &
  \multicolumn{1}{c|}{-} &
  - \\ \cline{3-10} 
 &
   &
  E3-Rewrite~\cite{xu2025e3_rewrite} &
  few-shot &
  \greencheck &
  \greencheck &
  \greencheck &
  \multicolumn{1}{c|}{-} &
  \multicolumn{1}{c|}{\greencheck} &
  - \\ \cline{2-10} 
 &
  \multirow{2}{*}{\begin{tabular}[c]{@{}c@{}}System \\ Diagnosis\end{tabular}} &
  DBG-PT~\cite{giannakouris2024dbg} &
  zero-shot &
  - &
  - &
  - &
  \multicolumn{1}{c|}{-} &
  \multicolumn{1}{c|}{\greencheck} &
  - \\ \cline{3-10} 
 &
   &
  Andromeda~\cite{chen2025andromeda} &
  few-shot &
  \greencheck &
  - &
  - &
  \multicolumn{1}{c|}{\greencheck} &
  \multicolumn{1}{c|}{\greencheck} &
  - \\ \hline
\multirow{11}{*}{\rotatebox{90}{\color[HTML]{000000}Data Preparation}} &
  \multirow{4}{*}{\begin{tabular}[c]{@{}c@{}}Data \\ Cleaning\end{tabular}} &
  FM~\cite{narayan2022can} &
  zero-shot &
  - &
  - &
  - &
  \multicolumn{1}{c|}{-} &
  \multicolumn{1}{c|}{-} &
  - \\ \cline{3-10} 
 &
   &
  RetClean~\cite{RetClean} &
  zero-shot &
  \greencheck &
  \greencheck &
  - &
  \multicolumn{1}{c|}{\greencheck} &
  \multicolumn{1}{c|}{\greencheck} &
  - \\ \cline{3-10} 
 &
   &
  LakeFill~\cite{lakefill} &
  zero-shot &
  \greencheck &
  \greencheck &
  - &
  \multicolumn{1}{c|}{\greencheck} &
  \multicolumn{1}{c|}{\greencheck} &
  - \\ \cline{3-10} 
 &
   &
  LLMClean~\cite{llmclean} &
  few-shot &
  - &
  - &
  - &
  \multicolumn{1}{c|}{\greencheck} &
  \multicolumn{1}{c|}{\greencheck} &
  - \\ \cline{2-10} 
 &
  \multirow{3}{*}{\begin{tabular}[c]{@{}c@{}}Data \\ Integration\end{tabular}} &
  Table-GPT~\cite{li2024table} &
  Chain-of-Thought &
  - &
  \greencheck &
  - &
  \multicolumn{1}{c|}{-} &
  \multicolumn{1}{c|}{-} &
  - \\ \cline{3-10} 
 &
   &
  BATCHER~\cite{fan2024cost} &
  Batch Prompting &
  - &
  - &
  - &
  \multicolumn{1}{c|}{-} &
  \multicolumn{1}{c|}{-} &
  - \\ \cline{3-10} 
 &
   &
  Jellyfish~\cite{zhang2024jellyfish} &
  few-shot/Chain-of-Thought &
  - &
  \greencheck &
  - &
  \multicolumn{1}{c|}{-} &
  \multicolumn{1}{c|}{-} &
  - \\ \cline{2-10} 
 &
  \multirow{4}{*}{\begin{tabular}[c]{@{}c@{}}Data \\ Discovery\end{tabular}} &
  ArcheType~\cite{feuer2024archetype}
 &
  zero-shot &
  - &
  \greencheck &
  - &
  \multicolumn{1}{c|}{\greencheck} &
  \multicolumn{1}{c|}{\greencheck} &
  - \\ \cline{3-10}  
 &
   &
  AutoDDG~\cite{zhang2025autoddg} &
  zero-shot &
  - &
  - &
  - &
  \multicolumn{1}{c|}{\greencheck} &
  \multicolumn{1}{c|}{\greencheck} &
  - \\ \cline{3-10} 
 &
   &
  Pneuma~\cite{balaka2025pneuma}
 &
  zero-shot &
  \greencheck &
  - &
  - &
  \multicolumn{1}{c|}{\greencheck} &
  \multicolumn{1}{c|}{\greencheck} &
  - \\ \cline{3-10} 
 &
   &
  LLMCTA~\cite{korini2025evaluating} &
  few-shot/Chain-of-Thought &
  - &
  \greencheck &
  - &
  \multicolumn{1}{c|}{\greencheck} &
  \multicolumn{1}{c|}{\greencheck} &
  - \\ \hline
\multirow{11}{*}{\rotatebox{90}{\color[HTML]{000000}Data Analysis}} &
  \multirow{4}{*}{\begin{tabular}[c]{@{}c@{}}Structured \\ Data \\ Analysis\end{tabular}} &
  BINDER~\cite{Binder2023} &
  few-shot &
  - &
  - &
  - &
  \multicolumn{1}{c|}{-} &
  \multicolumn{1}{c|}{\greencheck} &
  \greencheck \\ \cline{3-10} 
 &
   &
  DIN-SQL~\cite{dinsql} &
  few-shot &
  - &
  - &
  - &
  \multicolumn{1}{c|}{-} &
  \multicolumn{1}{c|}{-} &
  - \\ \cline{3-10} 
 &
   &
  Prompt4Vis~\cite{Prompt4Vis2025VLDB} &
  few-shot &
  - &
  - &
  - &
  \multicolumn{1}{c|}{\greencheck} &
  \multicolumn{1}{c|}{\greencheck} &
  - \\ \cline{3-10} 
 &
   &
  Step-Text2Vis~\cite{nvbench2-2025luo} &
  few-shot &
  - &
  \greencheck &
  \greencheck &
  \multicolumn{1}{c|}{-} &
  \multicolumn{1}{c|}{-} &
  - \\ \cline{2-10} 
 &
  \multirow{4}{*}{\begin{tabular}[c]{@{}c@{}}Unstructured \\ Data \\ Analysis\end{tabular}} &
  LongRAG~\cite{zhao2024longrag} &
  zero-shot &
  \greencheck &
  \greencheck &
  - &
  \multicolumn{1}{c|}{-} &
  \multicolumn{1}{c|}{-} &
  - \\ \cline{3-10} 
 &
   &
  PDFTriage~\cite{pdftriage2024} &
  zero-shot &
  \greencheck &
  - &
  - &
  \multicolumn{1}{c|}{-} &
  \multicolumn{1}{c|}{\greencheck} &
  - \\ \cline{3-10} 
 &
   &
  VisDoM~\cite{suri-etal-2025-visdom} &
  zero-shot &
  \greencheck &
  - &
  - &
  \multicolumn{1}{c|}{\greencheck} &
  \multicolumn{1}{c|}{\greencheck} &
  \greencheck \\ \cline{3-10} 
 &
   &
  Docopilot~\cite{docopilot} &
  zero-shot &
  - &
  \greencheck &
  - &
  \multicolumn{1}{c|}{\greencheck} &
  \multicolumn{1}{c|}{\greencheck} &
  \greencheck \\ \cline{2-10} 
 &
  \multirow{3}{*}{\begin{tabular}[c]{@{}c@{}}Report \\ Generation\end{tabular}} &
  Datatales~\cite{sultanum2023datatales} &few-shot &- &- &- &\multicolumn{1}{c|}{-} &\multicolumn{1}{c|}{-} &- \\ 
  \cline{3-10} 
 &
   &
  ReportGPT~\cite{cecchi2024reportgpt} & few-shot &- &- &- &\multicolumn{1}{c|}{-} &\multicolumn{1}{c|}{-} &- \\
  \cline{3-10} 
 &
   &
  ChartLens~\cite{suri-etal-2025-chartlens} &few-shot &- &- &- &\multicolumn{1}{c|}{-} &\multicolumn{1}{c|}{\greencheck} &
  \greencheck \\ 
  \hline
\end{tabular}} \vspace{-0em}
\end{table*}

Aligning with the early wave of LLM assistants, L1 data agents operate on a prompt-response basis, assisting with data-related tasks by providing organized answers or generating code snippets in response to user queries, as in Figure~\ref{fig:l1_agent}.

In this survey, we consider them nascent and underdeveloped data agents, which are yet stateless and lack environmental perception.
Formally, the human $\mathcal{H}$ remains responsible for both pipeline orchestration $\pi_{\mathcal{H}}$ and execution $\epsilon_{\mathcal{H}}$, while data agent $\mathcal{A}$ can respond $r$ upon human query $q$ for assistance:
\vspace{-.0em}
\begin{equation*}
\begin{aligned}
\mathcal{H} &: \pi_{\mathcal{H}}(\mathcal{T}, \mathcal{D}, \mathcal{E}) \rightarrow P; \quad \epsilon_{\mathcal{H}}(P, \mathcal{D}, \mathcal{E}, r) \rightarrow \mathcal{O} \\
\mathcal{A} &: (q, \mathcal{M}) \rightarrow r
\end{aligned}
\end{equation*}
\vspace{-.0em}

While L1 data agents enhance efficiency on routine tasks, their stateless nature limits them to generating immediate outputs without adaptive optimization based on environmental feedback, and humans still integrate, execute, and validate their output. This absence of environmental perception confines L1 data agents to isolated, one-off interactions, leaving end-to-end execution and verification under human control. Next, we discuss L1 data agents for data management, preparation, and analysis, with representative ones compared in Table~\ref{tab:l1_comparison}.

\vspace{-0em}
\subsection{L1 Assistance in Data Management}

Data management involves monitoring and optimizing database systems to ensure efficient and reliable operation. This section discusses L1 data agents in data management across three key tasks: (1) Configuration Tuning, (2) Query Optimization, and (3) System Diagnosis.

\stitle{Configuration Tuning.}
Configuration tuning aims to select and adjust database system settings like knobs and indexes to improve performance and efficiency~\cite{zhou2025survey}.
In the initial surge of LLMs, L1 data agents are developed to assist this by deploying LLMs to generate tuning recommendations from prompts customized with system information and domain knowledge.

\ptitle{Guiding Traditional Optimizers with LLMs.}
Earlier practices guide traditional optimization modules with LLMs.
LLMTune~\cite{huang2024llmtune} fine-tunes and prompts LLMs with workload information to generate a high-quality initial configuration, which a Bayesian Optimization (BO) optimizer then refines. Similarly, GPTuner~\cite{lao2024gptuner} refines domain knowledge via an LLM-driven pipeline and prompt ensembling to guide a coarse-to-fine BO framework. LATuner~\cite{fan2024latuner} constructs structured prompts for zero-shot warm-starting, using a hybrid LLM and Gaussian Process (GP) sampler with a multi-armed bandits-based adaptive surrogate model to select optimal configuration.

\ptitle{End-to-End LLM-Based Tuning.}
Some studies eliminate traditional optimizers, relying directly on LLM reasoning to overcome lengthy iterations, reconfiguration overhead, and transferability to unseen workloads.
For example, $\lambda$-Tune~\cite{giannakouris2025lambda} and E2ETune~\cite{huang2025e2etune} generate multiple configuration candidates based on workload features and use performance prediction guided by cost models to select the best option.

\stitle{Query Optimization.}
Query optimization seeks to improve the efficiency of SQL execution through logical and physical optimization. 
While traditional methods often neglect external SQL optimization knowledge, limiting their generalizability, L1 data agents leverage LLMs' encoded knowledge to determine rule sequences for logical optimization or select execution plans for physical optimization.

\etitle{Logical Query Optimization.}
DB-GPT~\cite{zhou2024db_gpt} incorporates an automatic prompt generation with adaptive demonstrations based on computed semantic similarity to enhance query rewriting. GenRewrite~\cite{liu2024genrewriter} introduces natural language rewrite rules as hints and a counterexample-guided CoT mechanism to self-correct rewritten queries. 
$\text{LLM-R}^2$~\cite{li2024llm_r2} recommends rewrite rules with a contrastive model trained via curriculum learning deployed as a few-shot demonstration manager. 
LITHE~\cite{dharwada2025lithe} combines a prompt ensemble with database-sensitive rules.
E3-Rewrite~\cite{xu2025e3_rewrite} advances by incorporating execution hints into prompts using parsed query plans and fine-tunes an LLM with a two-stage GRPO, prioritizing rewards of executability and semantic equivalence, leveraging execution plans to retrieve hybrid demonstrations during rewriting. 

\etitle{Physical Query Optimization.}
LLMOpt~\cite{yao2025llmopt} employs a fine-tuned LLM to generate multiple plan candidates and a separate LLM-based list-wise cost model to select the optimal query plan. LLM-QO~\cite{tan2025llmqo} constructs the QInstruct dataset, which sequentializes query plans, and applies DPO to train a policy model for generating more efficient execution plans.

\stitle{System Diagnosis.}
System diagnosis focuses on analyzing root causes and identifying recovery solutions for runtime anomalies in databases. L1 data agents start to demonstrate promising potential to pinpoint causes and generate diagnoses with recovery recommendations.

\ptitle{Comparative Diagnosis with LLMs.}
DBG-PT~\cite{giannakouris2024dbg} tackles unexpected slowdowns in query execution by enabling LLMs to compare regressed query plans with stored efficient instances, accordingly responding with configuration recommendations to resolve plan regressions. 

\ptitle{Augmenting Diagnosis with RAG.}
To avoid overly generic responses, Andromeda~\cite{chen2025andromeda}  incorporates a RAG mechanism to supply matched domain-specific context from historical queries, troubleshooting manuals, and DBMS telemetries, enhancing diagnostic suggestions for configuration debugging.

Despite advances, they rely on static inputs and lack real-time monitoring and interaction with databases, which limits their ability to provide adaptive diagnosis and recovery.

\vspace{-0em}
\subsection{L1 Assistance in Data Preparation}
Data preparation involves cleaning, integration, and discovery to ensure its quality and completeness for downstream tasks. This section reviews L1 data agents for this field.

\stitle{Data Cleaning.}
Data cleaning focuses on handling missing values, constraint violations, and inconsistencies within a dataset. With the emergence of LLMs, data agents at L1 are primarily employed in two modes for data cleaning:

\ptitle{Generating Direct Answers via Prompting.} 
A straightforward way is to directly prompt LLMs to infer and generate an answer (\eg filling a missing value)~\cite{foundation_wrangle_data}. For example, a data record can be serialized into text and fed to them to fill in a missing value with minimal examples~\cite{foundation_wrangle_data}. 
To improve quality, RetClean~\cite{RetClean} and LakeFill~\cite{lakefill} utilize an RAG approach, retrieving relevant tuples from the data lake to ground their response, which is crucial for private or domain-specific data. UniDM~\cite{qian2024unidm} employs a multi-stage prompting pipeline to guide the LLMs, identifying relevant metadata and instances, then parsing to construct targeted prompts.

\ptitle{Generating Intermediate Artifacts.} 
Alternatively, some systems generate intermediate artifacts like code or rules for the user to execute and validate.
For instance, LLMClean~\cite{llmclean} automates knowledge engineering by analyzing the given dataset to generate formal data quality rules, such as Ontological Functional Dependencies (OFDs), which are then passed to the user or a traditional cleaning system for implementation.

\stitle{Data Integration.}
Data integration refers to combining multi-source, heterogeneous data into a unified, coherent dataset~\cite{doan2012principles}, challenged by semantic heterogeneity, incomplete metadata, and entity ambiguity~\cite{freire2025large}. Data integration involves two key tasks: schema matching and entity resolution. 
Schema matching aligns data structures across different sources, while entity resolution builds on this alignment to resolve entity-level ambiguities, collectively achieving unification.

\etitle{Schema Matching.} 
The adoption of L1 data agents for schema matching has progressed from in-context learning to task-specific fine-tuning and integration into multi-stage pipelines, operating on a prompt-response basis essentially.

\ptitle{In-Context Learning and Fine-tuning.}
Some studies deploy off-the-shelf LLMs directly through zero-shot or few-shot learning. Narayan et al.~\cite{narayan2022can} serialize attribute values as input and showcase the effectiveness of few-shot learning in schema matching.
To improve adaptability, task-specific fine-tuning has been employed. Table-GPT~\cite{li2024table} performs multi-task fine-tuning, while Jellyfish~\cite{zhang2024jellyfish} focuses on meticulous prompt design and output reasoning.

\ptitle{Hybrid and Multi-Stage Systems.}
To address more complex matching scenarios, some approaches embed LLMs within multi-stage systems, which typically use pretrained embeddings to filter candidates, after which LLMs determine the final correspondences~\cite{seedat2024matchmaker, sheetrit2024rematch}. Magneto~\cite{liu2024magneto} exemplifies this by using a smaller language model (SLM) trained on synthesized data for initial ranking, followed by LLM re-ranking.

\etitle{Entity Resolution.} 
Narayan et al.~\cite{narayan2022can}  and Peeters et al.~\cite{peeters2023entity} use off-the-shelf LLMs with carefully designed prompts to produce entity matching decisions, presenting early exploration. Steiner et al.~\cite{steiner2025fine} optimize example selection and fine-tune LLMs with synthesized explanations on training data, improving in-domain generalization. Additionally, BatchER~\cite{fan2024cost} focuses on enhancing cost efficiency in in-context learning for entity resolution via batch prompting with a covering-based demonstration selection strategy.

\stitle{Data Discovery.}
Data discovery involves identifying, interpreting, and contextualizing datasets within large and heterogeneous repositories such as data lakes~\cite{freire2025large}. 
L1 data agents act as query responders, generating descriptive or structural artifacts (\eg summaries, annotations, profiles) upon human queries without altering the data.

\etitle{Descriptive Profiling.}  
A central role of these data agents is to act as descriptive interpreters that generate metadata or dataset profiles to facilitate accessibility. Within this role, existing approaches can be broadly divided into two categories: 

\ptitle{Prompt-Based Profiling.}
Representative work such as AutoDDG~\cite{zhang2025autoddg} integrates statistical and semantic information in prompts to create informative dataset summaries.
This paradigm is extended by others using domain-specific prompts for ecological metadata harvesting~\cite{lu2025flexible} or scientific paper validation~\cite{alyafeai2025mole}.
More advanced prompting strategies are used for semantic profiling, such as generating interpretable semantic views~\cite{huang2024cocoon} or customizable profiling for code data~\cite{thorat2025llm}.

\ptitle{Hybrid Retrieval Integration.}
The hybrid approach integrates retrieval to ground LLM outputs in repository-level signals, improving robustness and relevance. For example, Pneuma~\cite{balaka2025pneuma} combines LLM-based tabular representation with retrieval mechanisms for retrieval-aware profiling.

\etitle{Semantic Annotation for Schema.}  
Data agents are also deployed to serve as annotators, assigning semantic types, roles, or labels to schema elements.

\ptitle{Prompt-Based Annotation.}
Task-specific instructions within carefully crafted prompt templates are widely adopted at L1, and data agents directly infer semantics from column names or sampled values. 
ArcheType~\cite{feuer2024archetype} uses structured prompting and contextual sampling to guide LLMs in inferring column types.
LLMCTA~\cite{korini2025evaluating} extends to column property assignment, introducing knowledge-generation prompting to refine property definitions and assess annotation reliability across models. 
Moving further, Columbo~\cite{cai2025columbo} expands abbreviated or underspecified column names into semantically meaningful expressions, thereby enhancing schema interpretability. 

\ptitle{Semantic Grounding.}
To address inconsistency and robustness issues in prompt-based annotation, RACOON~\cite{wei2024racoon} leverages knowledge graphs to provide richer semantic grounding, while AutoMetaSQL~\cite{shkapenyuk2025automatic} mitigates context deficiency by extracting schema metadata directly from user queries. 

\vspace{-.0em}
\subsection{L1 Assistance in Data Analysis}

Data analysis involves processing and interpreting data to derive insights and support informed decision-making. This section reviews the application of L1 data agents across three areas: (1) Structured Data Analysis, (2) Unstructured Data Analysis, and (3) Report Generation.

\stitle{Structured Data Analysis.}
Structured data analysis centers on extracting insights from schema-organized data, such as tables or relational databases. In the following, we focus on three key tasks: TableQA, NL2SQL, and NL2VIS.

\etitle{TableQA} Table Question Answering (TableQA) enables interpreting natural language queries and extracting answers from structured or semi-structured tables~\cite{chen-etal-2020-hybridqa, zhu2024statqa, tang2026straptor}. It demands both semantic understanding of queries and parsing of the table's two-dimensional layout (\ie headers, rows, cells, and their relationships) to locate and infer responses.
Data agents for TableQA at L1 typically generate answers or code in a single-pass, static process, using various strategies.

\ptitle{Prompt Engineering.} 
Prompt engineering improves reasoning capabilities in TableQA, especially by mitigating the challenge of limited context windows for large tables via decomposition.
For instance, the Dater~\cite{Dater2023} uses an LLM to pre-process large tables into smaller, relevant ones and simplify questions. Similarly, self-augmented prompting~\cite{TableMeetsLLM2024} issues the first LLM call to extract structural insights that enrich the prompt for a second call to derive the final answer.

\ptitle{Fine-tuning.}
Instruction fine-tuning enhances LLMs' intrinsic reasoning capabilities for TableQA.
TableLlama~\cite{TableLlama2024} is fine-tuned on TableInstruct, a large-scale and diverse dataset of table-based instructions, enabling it to handle various table tasks directly without complex inference-time prompting.

\ptitle{Neuro-Symbolic Program Synthesis.}
A distinct methodology involves neuro-symbolic systems, which integrate LLMs' reasoning with the precision of formal code. Binder~\cite{Binder2023} exemplifies this by parsing natural language questions into hybrid programs that mix standard code (\eg SQL) with LLM API calls for common-sense reasoning, operating in a static ``parse-then-execute'' workflow.

\etitle{NL2SQL}
Natural Language to SQL (NL2SQL) converts a natural language query (NL) into a SQL query (SQL) to be executed on a relational database~\cite{liu2025survey, luo2025natural, liu2025nl2sql,supersql}, bridging non-technical users and databases.
Earlier efforts concentrated on the L1 autonomy primarily operating in a single-turn prompting paradigm~\cite{dailsql, dinsql, actsql}. 

\ptitle{Prompt Engineering.} 
DAIL-SQL~\cite{dailsql} systematically investigates prompt engineering for LLM-based NL2SQL, covering five forms of question representation, two prompt components, four example selection strategies, and three example organization schemes, with comprehensive evaluations conducted on four different LLMs. It also explores the effectiveness of supervised fine-tuning, extending the understanding of open-source models' applicability in NL2SQL.

\ptitle{Task Decomposition.}
DIN-SQL~\cite{dinsql} introduces a task decomposition-based prompting framework to address the limitations of single-turn zero/few-shot prompting. It splits SQL generation into sub-tasks (\eg schema linking, query sketching, clause completion), tackles step by step with crafted prompts, and progressively combines the results into a complete SQL, thereby improving performance on complex cases.

\ptitle{Self-Guided SQL Reasoning.}
ACT-SQL~\cite{actsql} focuses on enhancing the reasoning capability of LLMs in NL2SQL through automatic CoT prompting. Unlike approaches that rely on manually constructed exemplars, ACT-SQL automatically generates CoT sequences from the database schema, natural language queries, and corresponding SQL queries, effectively improving the accuracy of complex SQL generation.

\etitle{NL2VIS}
Natural Language to Visualization (NL2VIS) aims to enable users to create visualizations and derive analytical insights with natural language interfaces~\cite{ shen2023tvcg, nvBench2021, ncnet2022, visclean, linenet, deeptrack, DBLP:conf/icde/LuoQ0018}.
Leveraging LLMs' marked capabilities of understanding and reasoning, systems at L1 are prompted to provide on-demand assistance by translating users' intents in natural language into visualization specifications~\cite{Chat2VIS2023IEEE, xie2025visjudge}. 

\ptitle{Direct Prompting.}
Amid the rapid advancement of LLMs, Chat2VIS~\cite{Chat2VIS2023IEEE} is among the first to explore off-the-shelf LLMs to generate visualizations directly from natural language text. 
Prompt4Vis~\cite{Prompt4Vis2025VLDB} further enhances by employing automated prompt optimization, introducing a multi-objective example mining module to select the most effective examples for in-context learning. 
To address the ``black-box'' nature of LLMs, NL4DV-LLM~\cite{NL4DVLLM2024VIS} focuses on explainability by generating a structured analytic specification, which details inferred data attributes, tasks, and design rationale, making the process more transparent and easier to debug.

\ptitle{Step-Wise Reasoning.}
To address the persistent challenge of ambiguity of natural language queries in NL2VIS, nvBench 2.0~\cite{nvbench2-2025luo} introduces a benchmark with controlled ambiguity specifications using an ambiguity-injection pipeline and step-wise reasoning. It also presents Step-Text2Vis, which is trained to trace how each ambiguous query can result in multiple valid visualizations, setting a new state-of-the-art for ambiguity resolution in NL2VIS.

\stitle{Unstructured Data Analysis.}
Unstructured data analysis extracts insights from data lacking a predefined schema (\eg documents, images, etc.). Data agents at L1 have advanced this field through advanced semantic understanding and reasoning.

\etitle{Textual Documents.}
A primary challenge is comprehending vast and complex document corpora. Existing research has followed complementary paths to address this.

\ptitle{Intrinsic Context Expansion.}
One direction focuses on enhancing its intrinsic capacity to process vast, contiguous contexts~\cite{liu2025comprehensivesurveylongcontext}. As models with ever-larger context windows become available, techniques like R\&R~\cite{agrawal2024randr} have been developed to optimize how information is retrieved and utilized within these ultra-long inputs, significantly boosting accuracy. 

\ptitle{Structure-Aware Information Filtering.}
Concurrently, another strategy intelligently selects relevant information from the document, evolving beyond keyword retrieval to understand global document structure. RAPTOR~\cite{raptor2024} recursively clusters for hierarchical understanding, while LongRAG~\cite{zhao2024longrag} uses a dual-strategy for high-level summaries and fine-grained details. To handle complex layouts, PDFTriage~\cite{pdftriage2024} restructures documents into semantically coherent blocks.

\etitle{Multimodal Documents.}
Unstructured data analysis is broadening to complex multimodal formats integrating text, tables, and visuals, spurring relevant benchmarks like MMVQA~\cite{ding2024pdfmvqadatasetmultimodalinformation} and VisDoMBench~\cite{suri-etal-2025-visdom} for evaluation in visual-rich scenarios.

\ptitle{Multimodal RAG Adaptations.}
To tackle this, some studies adapt RAG. 
MMRAG-DocQA~\cite{gong2025mmrag} deploys hierarchical indexing and multi-granularity retrieval. A different paradigm, Document Screenshot Embedding~\cite{DSE2024}, bypasses traditional and error-prone content extraction by utilizing a vision-language model to directly encode visual appearance.

\ptitle{Multimodal Reasoning.}
Multimodal reasoning architectures also advance. VisDoM~\cite{suri-etal-2025-visdom} employs and fuses parallel textual and visual RAG pipelines by analyzing reasoning chain consistency. MDocAgent~\cite{han2025mdocagent} utilizes a multi-agent framework for collaboratively synthesizing cross-modal insights. Beyond RAG, end-to-end models like Docopilot~\cite{docopilot} perform multimodal understanding without an explicit retrieval component.

\vspace{-.0em}
\stitle{Report Generation.}
Report generation combines visualizations and narrative text to make analytical insights interpretable and actionable.
Within the early exploration of LLMs, data agents primarily act as interpreters and assistants by generating charts, summaries, or explanations in response to user queries, concentrating on the L1 autonomy.

Several representative works illustrate this pattern.
ReportGPT~\cite{cecchi2024reportgpt} introduces a verifiable table-to-text generation pipeline to derive human-readable narratives, which employs a DSL for explicit table operations and user feedback for factual verification. 
Datatales~\cite{sultanum2023datatales} investigates LLMs for authoring data-driven articles.
Similarly, InterChat~\cite{chen2025interchat}, VizTA~\cite{wang2025VizTA}, and Choe et al.~\cite{choe2025enhancing} explore agentic frameworks for multimodal interaction, using combined direct manipulation and natural language to disambiguate queries and enrich textual narratives with visually grounded evidence.
Extending this, ChartLens~\cite{suri-etal-2025-chartlens} introduces a multi-agent framework to ground narrative text with fine-grained visual attributes, providing bounding box citations as evidence within charts.

\vspace{-.0em}
\subsection{Progress and Limitations of L1 Data Agents}
L1 data agents have introduced prompt-based query-responsive assistance for discrete data-related tasks. 
By interpreting and responding to users' queries on demand, this initial form of intelligent assistance also significantly improves developer efficiency by offloading trivial and routine operations.

As summarized in Table~\ref{tab:l1_comparison}, representative L1 data agents mainly enhance prompt-response assistance through in-context learning, task-specific prompting, RAG, SFT, and occasionally RL. These methods improve the relevance and quality of generated answers, rules, code snippets, or descriptions for specific data-related subtasks, while leaving environmental feedback and execution responsibility outside the data agent.

Accordingly, L1 data agents still operate in a stateless, prompt–response paradigm and lack the capabilities to perceive and interact with their environment, limiting their ability to perform adaptive optimization. They can not perceive or interact with external environments or data systems autonomously, necessitating manual execution, integration, verification, and optimization of generated outputs by human analysts. Consequently, L1 data agents remain reliant on user oversight to ensure correctness and to manage the entire procedure, limiting their autonomy to atomic and static subtasks.

\section{L2: Perceive the Environment}
\label{sec:l2}

\begin{figure}[t!]
    \centering
    \includegraphics[width=\linewidth]{\paperroot 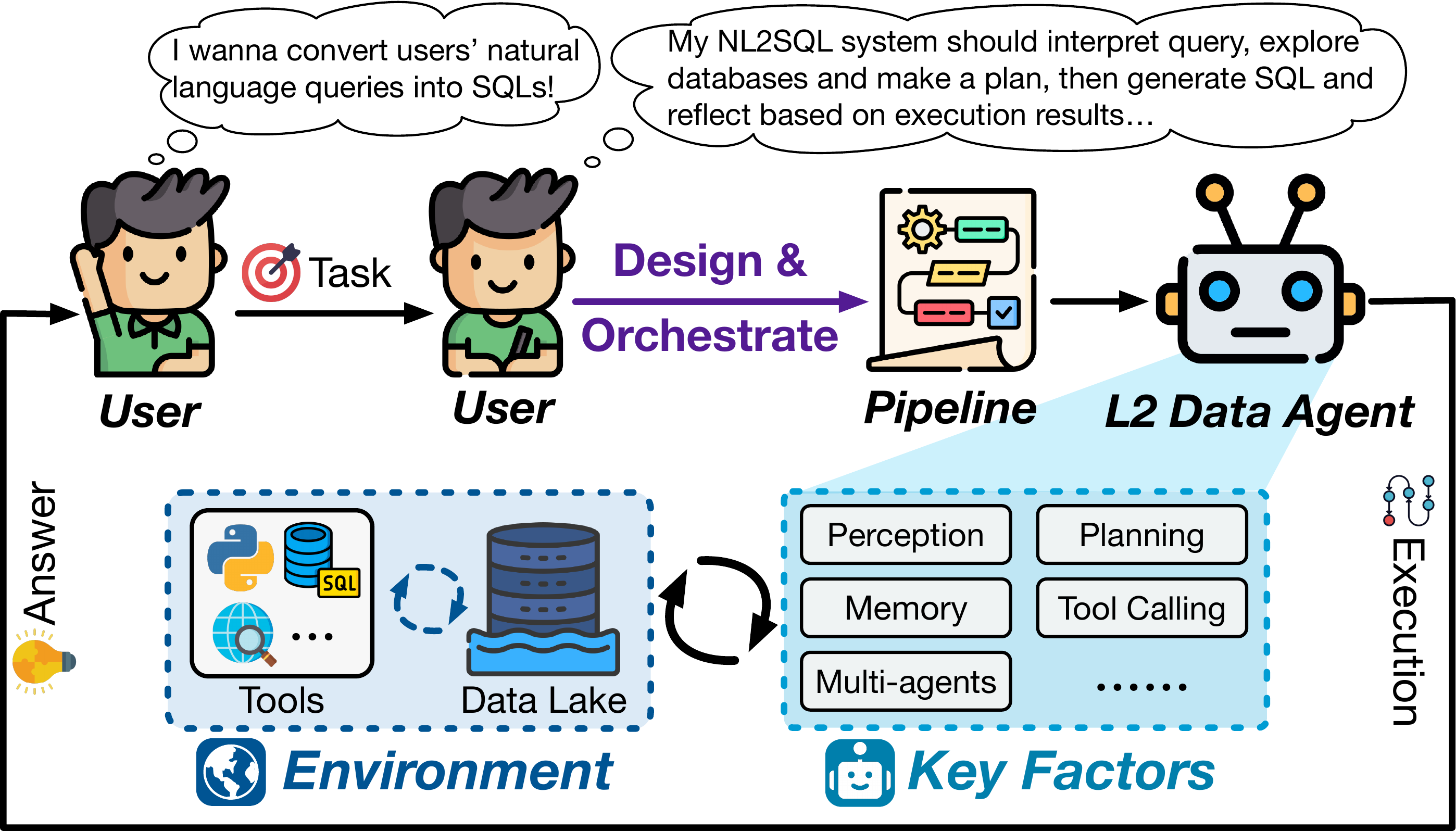}
    \vspace{-0em}
    \caption{L2 Data Agents (Partial Autonomy).}
    \label{fig:l2_agent}
    \vspace{-0em}
\end{figure}

\begingroup
\renewcommand{\arraystretch}{0.99}
\begin{table*}[t!]
\centering
\caption{Comparison of Representative L2 Data Agents. RAG: Retrieval-Augmented Generation; Percept: Environmental Perception; Plan: Planning; Mem: Memory; Tool: Tool Invocation; Reflect: Self-reflection mechanism; MAS: Multi-agent system; SFT: Supervised Fine-Tuning; RL: Reinforcement Learning. Data complexity: Multi-source (Multis.), Heterogeneous (Hete.), and Multimodal (Multim.).}
\vspace{-0em}
\label{tab:l2_comparison}
\resizebox{\textwidth}{!}{\begin{tabular}{|c|c|c|c|c|c|c|c|c|c|c|c|ccc|}
\hline
\multirow{2}{*}{\textbf{}} &
  \multirow{2}{*}{\textbf{Task}} &
  \multirow{2}{*}{\textbf{Data Agent}} &
  \multirow{2}{*}{\textbf{RAG}} &
  \multirow{2}{*}{\textbf{Percept}} &
  \multirow{2}{*}{\textbf{Plan}} &
  \multirow{2}{*}{\textbf{Mem}} &
  \multirow{2}{*}{\textbf{Tool}} &
  \multirow{2}{*}{\textbf{Reflect}} &
  \multirow{2}{*}{\textbf{MAS}} &
  \multirow{2}{*}{\textbf{SFT}} &
  \multirow{2}{*}{\textbf{RL}} &
  \multicolumn{3}{c|}{\textbf{Data Complexity}} \\ \cline{13-15} 
 &
   &
   &
   &
   &
   &
   &
   &
   &
   &
   &
   &
  \multicolumn{1}{c|}{\textbf{Multis.}} &
  \multicolumn{1}{c|}{\textbf{Hete.}} &
  \textbf{Multim.} \\ \hline
\multirow{10}{*}{\rotatebox{90}{\color[HTML]{000000}Data Management}} &
  \multirow{4}{*}{\begin{tabular}[c]{@{}c@{}}Configuration \\ Tuning\end{tabular}} &
  Li et al.~\cite{li2024large} &
  - &
  \greencheck &
  - &
  \greencheck &
  - &
  - &
  - &
  - &
  - &
  \multicolumn{1}{c|}{-} &
  \multicolumn{1}{c|}{-} &
  - \\ \cline{3-15} 
 &
   &
  LLMIdxAdvis~\cite{zhao2025llmidxadvis} &
  \greencheck &
  \greencheck &
  - &
  \greencheck &
  - &
  \greencheck &
  \greencheck &
  - &
  - &
  \multicolumn{1}{c|}{-} &
  \multicolumn{1}{c|}{\greencheck} &
  - \\ \cline{3-15} 
 &
   &
  RABBIT~\cite{sun2025rabbit} &
  \greencheck &
  \greencheck &
  - &
  - &
  - &
  \greencheck &
  \greencheck &
  - &
  - &
  \multicolumn{1}{c|}{\greencheck} &
  \multicolumn{1}{c|}{\greencheck} &
  - \\ \cline{3-15} 
 &
   &
  MCTuner~\cite{yan2025mctuner} &
  \greencheck &
  \greencheck &
  - &
  \greencheck &
  \greencheck &
  - &
  \greencheck &
  - &
  - &
  \multicolumn{1}{c|}{\greencheck} &
  \multicolumn{1}{c|}{\greencheck} &
  - \\ \cline{2-15} 
 &
  \multirow{4}{*}{\begin{tabular}[c]{@{}c@{}}Query \\ Optimization\end{tabular}} &
  SERAG~\cite{liu2025serag} &
  \greencheck &
  \greencheck &
  - &
  \greencheck &
  - &
  - &
  - &
  - &
  - &
  \multicolumn{1}{c|}{-} &
  \multicolumn{1}{c|}{-} &
  - \\ \cline{3-15} 
 &
   &
  QUITE~\cite{song2025quite} &
  \greencheck &
  \greencheck &
  \greencheck &
  \greencheck &
  \greencheck &
  \greencheck &
  \greencheck &
  - &
  - &
  \multicolumn{1}{c|}{\greencheck} &
  \multicolumn{1}{c|}{-} &
  - \\ \cline{3-15} 
  &
   &
  R-Bot~\cite{sun2025rbot} &
  \greencheck &
  \greencheck &
  - &
  - &
  \greencheck &
  \greencheck &
  - &
  - &
  - &
  \multicolumn{1}{c|}{\greencheck} &
  \multicolumn{1}{c|}{\greencheck} &
  - \\ \cline{3-15} 
 &
   &
  CrackSQL~\cite{cracksql} &
  \greencheck &
  \greencheck &
  - &
  - &
  - &
  \greencheck &
  - &
  \greencheck &
  - &
  \multicolumn{1}{c|}{\greencheck} &
  \multicolumn{1}{c|}{\greencheck} &
  - \\ \cline{2-15} 
 &
  \multirow{3}{*}{\begin{tabular}[c]{@{}c@{}}System \\ Diagnosis\end{tabular}} &
  Panda~\cite{singh2024panda} &
  \greencheck &
  \greencheck &
  - &
  \greencheck &
  - &
  - &
  - &
  - &
  - &
  \multicolumn{1}{c|}{\greencheck} &
  \multicolumn{1}{c|}{\greencheck} &
  \greencheck \\ \cline{3-15} 
 &
   &
  D-Bot~\cite{zhou2024dbot} &
  \greencheck &
  \greencheck &
  \greencheck &
  \greencheck &
  \greencheck &
  \greencheck &
  \greencheck &
  \greencheck &
  - &
  \multicolumn{1}{c|}{\greencheck} &
  \multicolumn{1}{c|}{\greencheck} &
  - \\ \cline{3-15} 
 &
   &
  DBAIOps~\cite{zhou2025dbaiops} &
  \greencheck &
  \greencheck &
  - &
  - &
  \greencheck &
  - &
  - &
  - &
  - &
  \multicolumn{1}{c|}{\greencheck} &
  \multicolumn{1}{c|}{\greencheck} &
  - \\ \hline
\multirow{17}{*}{\rotatebox{90}{\color[HTML]{000000}Data Preparation}} &
  \multirow{7}{*}{\begin{tabular}[c]{@{}c@{}}Data \\ Cleaning\end{tabular}} &
  SketchFill~\cite{zhang2024sketchfill} &
  - &
  \greencheck &
  - &
  - &
  \greencheck &
  \greencheck &
  \greencheck &
  - &
  - &
  \multicolumn{1}{c|}{-} &
  \multicolumn{1}{c|}{-} &
  - \\ \cline{3-15} 
 &
   &
  IterClean~\cite{IterClean} &
  - &
  \greencheck &
  - &
  - &
  \greencheck &
  \greencheck &
  \greencheck &
  - &
  - &
  \multicolumn{1}{c|}{-} &
  \multicolumn{1}{c|}{-} &
  - \\ \cline{3-15} 
   &
   &
  AutoPrep~\cite{fan2025autoprep} &
  - &
  \greencheck &
  \greencheck &
  - &
  \greencheck &
  - &
  \greencheck &
  - &
  - &
  \multicolumn{1}{c|}{-} &
  \multicolumn{1}{c|}{\greencheck} &
  - \\ \cline{3-15} 
 &
   &
  AutoDCWorkflow~\cite{li2025autodcworkflow} &
  - &
  \greencheck &
  \greencheck &
  - &
  \greencheck &
  \greencheck &
  - &
  - &
  - &
  \multicolumn{1}{c|}{\greencheck} &
  \multicolumn{1}{c|}{\greencheck} &
  - \\ \cline{3-15} 
 &
   &
  CleanAgent~\cite{qi2025cleanagentautomatingdatastandardization} &
  - &
  \greencheck &
  \greencheck &
  \greencheck &
  \greencheck &
  - &
  \greencheck &
  - &
  - &
  \multicolumn{1}{c|}{-} &
  \multicolumn{1}{c|}{\greencheck} &
  - \\ \cline{3-15} 
 &
   &
  Bendinelli et al.~\cite{bendinelli2025exploring} &
  - &
  \greencheck &
  \greencheck &
  - &
  \greencheck &
  - &
  - &
  - &
  - &
  \multicolumn{1}{c|}{-} &
  \multicolumn{1}{c|}{-} &
  - \\ \cline{3-15} 
 &
   &
  MegaTran~\cite{li2025megatran} &
  \greencheck &
  \greencheck &
  - &
  - &
  \greencheck &
  \greencheck &
  - &
  \greencheck &
  - &
  \multicolumn{1}{c|}{\greencheck} &
  \multicolumn{1}{c|}{\greencheck} &
  - \\ \cline{2-15} 
 &
  \multirow{3}{*}{\begin{tabular}[c]{@{}c@{}}Data \\ Integration\end{tabular}} 
     &
  Agent-OM~\cite{qiang2024agent} &
  \greencheck &
  \greencheck &
  \greencheck &
  - &
  \greencheck &
  \greencheck &
  \greencheck &
  - &
  - &
  \multicolumn{1}{c|}{\greencheck} &
  \multicolumn{1}{c|}{-} &
  - \\ \cline{3-15} 
 &
  &
  MILA~\cite{taboada2025ontology} &
  \greencheck &
  \greencheck &
  \greencheck &
  - &
  - &
  - &
  - &
  - &
  - &
  \multicolumn{1}{c|}{\greencheck} &
  \multicolumn{1}{c|}{-} &
  - \\ \cline{3-15} 
 &
   &
  COMEM~\cite{wang2025match} &
  - &
  \greencheck &
  \greencheck &
  - &
  - &
  - &
  - &
  - &
  - &
  \multicolumn{1}{c|}{\greencheck} &
  \multicolumn{1}{c|}{-} &
  - \\ \cline{2-15} 
 &
  \multirow{7}{*}{\begin{tabular}[c]{@{}c@{}}Data \\ Discovery\end{tabular}} &
  Chorus~\cite{kayali2024chorus} &
  - &
  \greencheck &
  - &
  - &
  - &
  - &
  \greencheck &
  - &
  - &
  \multicolumn{1}{c|}{-} &
  \multicolumn{1}{c|}{\greencheck} &
  - \\ \cline{3-15} 
 &
   &
  DataVoyager~\cite{majumder2024data} &
  - &
  \greencheck &
  \greencheck &
  \greencheck &
  \greencheck &
  \greencheck &
  \greencheck &
  - &
  - &
  \multicolumn{1}{c|}{\greencheck} &
  \multicolumn{1}{c|}{\greencheck} &
  \greencheck \\ \cline{3-15} 
 &
   &
  DATALORE~\cite{lou2024datalore} &
  - &
  \greencheck &
  \greencheck &
  - &
  \greencheck &
  \greencheck &
  - &
  - &
  - &
  \multicolumn{1}{c|}{\greencheck} &
  \multicolumn{1}{c|}{-} &
  - \\ \cline{3-15} 
 &
   &
  LEDD~\cite{an2025ledd} &
  - &
  \greencheck &
  \greencheck &
  - &
  \greencheck &
  \greencheck &
  - &
  - &
  - &
  \multicolumn{1}{c|}{\greencheck} &
  \multicolumn{1}{c|}{-} &
  - \\ \cline{3-15} 
 &
   &
  MetaGen~\cite{alrubaye2025metadata} &
  - &
  \greencheck &
  - &
  - &
  - &
  - &
  - &
  - &
  - &
  \multicolumn{1}{c|}{\greencheck} &
  \multicolumn{1}{c|}{-} &
  - \\ \cline{3-15} 
 &
   &
  DBDescGen~\cite{li2025autodb} &
  - &
  \greencheck &
  \greencheck &
  \greencheck &
  \greencheck &
  \greencheck &
  \greencheck &
  - &
  - &
  \multicolumn{1}{c|}{\greencheck} &
  \multicolumn{1}{c|}{\greencheck} &
  - \\ \cline{3-15} 
 &
   &
  Wang et al.~\cite{wang2025towards} &
  - &
  \greencheck &
  \greencheck &
  - &
  \greencheck &
  \greencheck &
  \greencheck &
  - &
  - &
  \multicolumn{1}{c|}{\greencheck} &
  \multicolumn{1}{c|}{\greencheck} &
  \greencheck \\ \hline
\multirow{33}{*}{\rotatebox{90}{\color[HTML]{000000}Data Analysis}} &
  \multirow{17}{*}{\begin{tabular}[c]{@{}c@{}}Structured \\ Data \\ Analysis\end{tabular}} &
  StructGPT~\cite{StructGPT2023} &
  - &
  \greencheck &
  \greencheck &
  - &
  \greencheck &
  - &
  - &
  - &
  - &
  \multicolumn{1}{c|}{\greencheck} &
  \multicolumn{1}{c|}{\greencheck} &
  - \\ \cline{3-15} 
 &
  &
  ReAcTable~\cite{ReAcTable2024} &
  - &
  \greencheck &
  \greencheck &
  - &
  \greencheck &
  - &
  - &
  - &
  - &
  \multicolumn{1}{c|}{-} &
  \multicolumn{1}{c|}{-} &
  - \\ \cline{3-15} 
 &
   &
  Chain-of-Table~\cite{CHAINOFTABLE2024} &
  - &
  \greencheck &
  \greencheck &
  - &
  \greencheck &
  - &
  - &
  - &
  - &
  \multicolumn{1}{c|}{-} &
  \multicolumn{1}{c|}{-} &
  - \\ \cline{3-15} 
 &
   &
  AutoTQA~\cite{AutoTQA2024} &
  - &
  \greencheck &
  \greencheck &
  - &
  \greencheck &
  \greencheck &
  \greencheck &
  - &
  - &
  \multicolumn{1}{c|}{\greencheck} &
  \multicolumn{1}{c|}{\greencheck} &
  - \\ \cline{3-15} 
 &
   &
  Table-Critic~\cite{TableCritic2025} &
  - &
  \greencheck &
  - &
  \greencheck &
  - &
  \greencheck &
  \greencheck &
  - &
  - &
  \multicolumn{1}{c|}{-} &
  \multicolumn{1}{c|}{-} &
  - \\ \cline{3-15} 
 &
   &
  ST-Raptor~\cite{tang2026straptor} &
  - &
  \greencheck &
  \greencheck &
  \greencheck &
  - &
  - &
  \greencheck &
  - &
  - &
  \multicolumn{1}{c|}{-} &
  \multicolumn{1}{c|}{-} &
  - \\ \cline{3-15} 
 &
  &
  MAC-SQL~\cite{macsql} &
  - &
  \greencheck &
  \greencheck &
  - &
  \greencheck &
  \greencheck &
  \greencheck &
  - &
  - &
  \multicolumn{1}{c|}{-} &
  \multicolumn{1}{c|}{-} &
  - \\ \cline{3-15} 
 &
     &
      ChatBI~\cite{lian2024chatbi} &
      - &
      \greencheck &
      \greencheck &
      \greencheck &
      \greencheck &
      \greencheck &
      - &
      - &
      - &
      \multicolumn{1}{c|}{\greencheck} &
      \multicolumn{1}{c|}{\greencheck} &
      - \\ \cline{3-15} 
     &
    &
  Chase-SQL~\cite{pourreza2025chasesql} &
  - &
  \greencheck &
  \greencheck &
  - &
  \greencheck &
  \greencheck &
  \greencheck &
  \greencheck &
  - &
  \multicolumn{1}{c|}{\greencheck} &
  \multicolumn{1}{c|}{\greencheck} &
  - \\ \cline{3-15} 
 &
    &
  Alpha-SQL~\cite{li2025alphasql} &
  - &
  \greencheck &
  - &
  - &
  \greencheck &
  \greencheck &
  \greencheck &
  - &
  - &
  \multicolumn{1}{c|}{-} &
  \multicolumn{1}{c|}{-} &
  - \\ \cline{3-15} 
 &
   &
  OpenSearch-SQL~\cite{opensearchsql} &
  \greencheck &
  \greencheck &
  - &
  - &
  \greencheck &
  \greencheck &
  \greencheck &
  - &
  - &
  \multicolumn{1}{c|}{\greencheck} &
  \multicolumn{1}{c|}{\greencheck} &
  - \\ \cline{3-15} 
 &
   &
  ReFoRCE~\cite{deng2025reforce} &
  - &
  \greencheck &
  - &
  \greencheck &
  \greencheck &
  \greencheck &
  \greencheck &
  - &
  - &
  \multicolumn{1}{c|}{-} &
  \multicolumn{1}{c|}{\greencheck} &
  - \\ \cline{3-15} 
  &
   & 
 DeepEye-SQL~\cite{li2025deepeyesql} &
  \greencheck &
  \greencheck &
  - &
  - &
  \greencheck &
  \greencheck &
  \greencheck &
  - &
  - &
  \multicolumn{1}{c|}{\greencheck} &
  \multicolumn{1}{c|}{\greencheck} &
  - \\ \cline{3-15} 
 &
  &
  MatPlotAgent~\cite{MatPlotAgent2024ACL} &
  - &
  \greencheck &
  - &
  - &
  \greencheck &
  \greencheck &
  \greencheck &
  - &
  - &
  \multicolumn{1}{c|}{-} &
  \multicolumn{1}{c|}{-} &
  - \\ \cline{3-15} 
 &
   &
  Text2Chart31~\cite{Text2Chart312024EMNLP} &
  - &
  \greencheck &
  - &
  - &
  \greencheck &
  \greencheck &
  - &
  \greencheck &
  - &
  \multicolumn{1}{c|}{\greencheck} &
  \multicolumn{1}{c|}{\greencheck} &
  - \\ \cline{3-15} 
 &
   &
  nvAgent~\cite{nvAgent2025ACL} &
  - &
  \greencheck &
  \greencheck &
  - &
  \greencheck &
  \greencheck &
  \greencheck &
  - &
  - &
  \multicolumn{1}{c|}{\greencheck} &
  \multicolumn{1}{c|}{\greencheck} &
  - \\ \cline{3-15} 
 &
   &
  DeepVIS~\cite{shuai2025deepvis} &
  - &
  \greencheck &
  \greencheck &
  - &
  \greencheck &
  \greencheck &
  - &
  \greencheck &
  - &
  \multicolumn{1}{c|}{-} &
  \multicolumn{1}{c|}{-} &
  - \\ \cline{2-15} 
 &
  \multirow{9}{*}{\begin{tabular}[c]{@{}c@{}}Unstructured \\ Data \\ Analysis\end{tabular}} &
  FLARE~\cite{flare2023} &
  \greencheck &
  \greencheck &
  \greencheck &
  - &
  \greencheck &
  - &
  - &
  - &
  - &
  \multicolumn{1}{c|}{\greencheck} &
  \multicolumn{1}{c|}{-} &
  - \\ \cline{3-15}
 &
   &
  ReadAgent~\cite{readagent2024} &
  - &
  \greencheck &
  \greencheck &
  \greencheck &
  - &
  - &
  - &
  - &
  - &
  \multicolumn{1}{c|}{-} &
  \multicolumn{1}{c|}{-} &
  - \\ \cline{3-15} 
 &
   &
  GraphReader~\cite{graphreader2024} &
  - &
  \greencheck &
  \greencheck &
  \greencheck &
  \greencheck &
  \greencheck &
  - &
  - &
  - &
  \multicolumn{1}{c|}{-} &
  \multicolumn{1}{c|}{-} &
  - \\ \cline{3-15} 
 &
   &
  Self-RAG~\cite{asai2023selfrag} &
  \greencheck &
  \greencheck &
  \greencheck &
  - &
  \greencheck &
  \greencheck &
  - &
  \greencheck &
  - &
  \multicolumn{1}{c|}{\greencheck} &
  \multicolumn{1}{c|}{\greencheck} &
  - \\ \cline{3-15} 
 &
   &
  REAR~\cite{rear2024} &
  \greencheck &
  \greencheck &
  - &
  - &
  - &
  - &
  - &
  \greencheck &
  - &
  \multicolumn{1}{c|}{-} &
  \multicolumn{1}{c|}{-} &
  - \\ \cline{3-15} 
 &
   &
  QUEST~\cite{sun2025quest} &
  \greencheck &
  \greencheck &
  \greencheck &
  - &
  \greencheck &
  - &
  - &
  - &
  - &
  \multicolumn{1}{c|}{\greencheck} &
  \multicolumn{1}{c|}{\greencheck} &
  - \\ \cline{3-15} 
 &
   &
  Doctopus~\cite{chai2025doctopus} &
  \greencheck &
  \greencheck &
  \greencheck &
  - &
  \greencheck &
  - &
  - &
  \greencheck &
  - &
  \multicolumn{1}{c|}{\greencheck} &
  \multicolumn{1}{c|}{\greencheck} &
  - \\ \cline{3-15} 
 &
   &
  MACT~\cite{yu2025mact} &
  - &
  \greencheck &
  \greencheck &
  \greencheck &
  - &
  \greencheck &
  \greencheck &
  \greencheck &
  \greencheck &
  \multicolumn{1}{c|}{\greencheck} &
  \multicolumn{1}{c|}{\greencheck} &
  \greencheck \\ \cline{3-15} 
 &
   &
  DataPuzzle~\cite{zhang2025datapuzzle} &
  \greencheck &
  \greencheck &
  \greencheck &
  - &
  - &
  \greencheck &
  \greencheck &
  - &
  - &
  \multicolumn{1}{c|}{\greencheck} &
  \multicolumn{1}{c|}{\greencheck} &
  \greencheck \\ \cline{2-15} 
 &
  \multirow{10}{*}{\begin{tabular}[c]{@{}c@{}}Report \\ Generation\end{tabular}} &
  DataNarrative~\cite{islam2024datanarrative} &
  - &
  \greencheck &
  - &
  - &
  \greencheck &
  \greencheck &
  \greencheck &
  - &
  - &
  \multicolumn{1}{c|}{-} &
  \multicolumn{1}{c|}{\greencheck} &
  - \\ \cline{3-15} 
 &
   &
  Data Director~\cite{shen2024datadirector} &
  - &
  \greencheck &
  \greencheck &
  - &
  \greencheck &
  - &
  \greencheck &
  - &
  - &
  \multicolumn{1}{c|}{-} &
  \multicolumn{1}{c|}{-} &
  - \\ \cline{3-15}
 &
   &
  DataMagic~\cite{xie2026datamagic} &
  - &
  \greencheck &
  \greencheck &
  - &
  \greencheck &
  - &
  \greencheck &
  - &
  - &
  \multicolumn{1}{c|}{-} &
  \multicolumn{1}{c|}{-} &
  - \\ \cline{3-15}
  &
   &
  LightVA~\cite{zhao2025lightva} &
  - &
  \greencheck &
  \greencheck &
  \greencheck &
  \greencheck &
  \greencheck &
  \greencheck &
  - &
  - &
   \multicolumn{1}{c|}{\greencheck} &
  \multicolumn{1}{c|}{\greencheck} &
  - \\ \cline{3-15} 
 &
   &
  \begin{tabular}[c]{@{}c@{}}Multimodal \\[-3pt] DeepResearcher~\cite{yang2025deepresearcher}\end{tabular} &
  \greencheck &
  \greencheck &
  - &
  - &
  \greencheck &
  \greencheck &
  \greencheck &
  - &
  - &
  \multicolumn{1}{c|}{\greencheck} &
  \multicolumn{1}{c|}{\greencheck} &
  \greencheck \\ \cline{3-15} 
 &
   &
  DAgent~\cite{xu2025dagent} &
  \greencheck &
  \greencheck &
  \greencheck &
  \greencheck &
  \greencheck &
  - &
  - &
  \greencheck &
  - &
  \multicolumn{1}{c|}{-} &
  \multicolumn{1}{c|}{\greencheck} &
  - \\ \cline{3-15} 
 &
   &
  Chart Insighter~\cite{wang2025chartinsighter} &
  - &
  \greencheck &
  \greencheck &
  - &
  \greencheck &
  \greencheck &
  \greencheck &
  - &
  - &
  \multicolumn{1}{c|}{-} &
  \multicolumn{1}{c|}{\greencheck} &
  - \\ \cline{3-15} 
 &
   &
  ProactiveVA~\cite{zhao2025proactiveva} &
  - &
  \greencheck &
  \greencheck &
  \greencheck &
  \greencheck &
  \greencheck &
  \greencheck &
  - &
  - &
  \multicolumn{1}{c|}{\greencheck} &
  \multicolumn{1}{c|}{\greencheck} &
  \greencheck \\ \cline{3-15} 
 &
   &
  VOICE~\cite{jia2024voice} &
  - &
  \greencheck &
  \greencheck &
  \greencheck &
  \greencheck &
  \greencheck &
  \greencheck &
  \greencheck &
  - &
  \multicolumn{1}{c|}{\greencheck} &
  \multicolumn{1}{c|}{\greencheck} &
  \greencheck \\ \cline{3-15} 
 &
   &
  NLI4VolVis~\cite{ai2025nli4volvis} &
  - &
  \greencheck &
  \greencheck &
  \greencheck &
  \greencheck &
  \greencheck &
  \greencheck &
  - &
  - &
  \multicolumn{1}{c|}{\greencheck} &
  \multicolumn{1}{c|}{\greencheck} &
  \greencheck \\ \hline

\end{tabular}}
\vspace{-0em}
\end{table*}
\endgroup

\subsection{Partial Autonomy: From Responder to Executor}
\vspace{-0em}

As outlined in Section~\ref{sec:levels}, L2 data agents gain the ability to perceive and interact with their environment, evolving from query-responsive assistance to procedural executors within human-designed pipelines. As illustrated in Figure~\ref{fig:l2_agent}, unlike L1 where humans are responsible for interacting with the environment, L2 data agents connect to real-world data systems such as DBMS, files, and APIs, enabling them to perceive and explore these sources, execute operations, invoke external tools (\eg SQL equivalence checker, code sandboxes, visualization libraries etc.), and adaptively optimize outputs based on environmental feedback (\eg query results or execution outcomes). This grants L2 data agents partial autonomy to perform task-specific procedures independently.
Formally, at L2, data agent $\mathcal{A}$ gains environmental perception and interaction capabilities ($\mathcal{D, E}$), capable of handling specific data-related tasks by executing ($\epsilon_{\mathcal{A}}$) pipeline $P$ orchestrated by human $\mathcal{H}$:
\vspace{-.0em}
\begin{equation*}
\begin{aligned}
\mathcal{H} &: \pi_{\mathcal{H}}(\mathcal{T}, \mathcal{D}, \mathcal{E}) \rightarrow P \\
\mathcal{A} &: \epsilon_{\mathcal{A}}(P, \mathcal{D}, \mathcal{E}, \mathcal{M}) \rightarrow \mathcal{O}
\end{aligned}
\end{equation*}
\vspace{-.0em}

This represents a significant advancement beyond the stateless prompt-response paradigm at L1, alleviating users from the repetitive cycle of integrating data agents' outputs and interacting with the environment manually. In practice, many recent developments fall within the L2. For instance, ReFoRCE~\cite{deng2025reforce} actively explores database columns and translates and optimizes natural language queries into SQL through interaction with the database. However, the autonomy of L2 data agents remains limited: they operate within human-orchestrated pipelines and are not yet capable of independently devising workflows tailored to diverse, comprehensive tasks. 

Next, we review L2 data agents for various data-related tasks with Table~\ref{tab:l2_comparison} outlining a summary and comparison.

\vspace{-0em}
\subsection{L2 Data Agents in Data Management}
\vspace{-0em}

\stitle{Configuration Tuning.}
L2 data agents not only generate configuration recommendations but also monitor system states to dynamically refine tuning strategies based on real-time performance feedback. 
They continuously adjust parameters such as database knobs, using either direct system feedback or structured knowledge sources to improve effectiveness.

\ptitle{Iterative Tuning via Feedback.}
A typical approach is direct feedback-driven interaction, where data agents evaluate and refine tuning suggestions using execution metrics from database systems.
Li et al.~\cite{li2024large} introduce a feedback loop for LLMs to iteratively manage knob selection, initialization, and recommendation based on performance metric outcomes. 
LLMIdxAdvis~\cite{zhao2025llmidxadvis} employs a similar self-optimization mechanism for iterative index tuning, using feedback from estimated storage and cost reduction to guide refinements.

\ptitle{Knowledge-Guided Tuning.}
Other studies further use structured domain knowledge from knowledge bases (\eg knowledge graphs or expert repositories) to guide and validate tuning actions. 
Rabbit~\cite{sun2025rabbit} retrieves knowledge from a knob-centric knowledge graph that encodes static database manuals, knob dependencies, and historical tuning experiences, accordingly adjusting search strategy using similarity-based feedback from past tasks. 
MCTuner~\cite{yan2025mctuner} employs a Mixture-of-Experts (MoE) mechanism to identify performance-critical knobs and validating multi-source knowledge to guide selection under multi-objective constraints.

\ptitle{Limitations.}
L2 data agents can perceive system states and invoke tools to recommend configurations, but remain confined to pre-defined knob sets and procedures within a guided pipeline, lacking autonomy to self-initiate tuning processes or adapt settings dynamically without human intervention.

\stitle{Query Optimization.}
L2 data agents for query optimization employ LLMs as reasoning engines to iteratively explore alternative query formulations or execution plans, incorporating database feedback or external tools to enhance efficiency.

\ptitle{Execution Feedback-Driven Refinement.}
Direct approaches focus on leveraging real query execution feedback for iterative refinement of optimization strategies.
SERAG~\cite{liu2025serag} retrieves past examples in a vector database to generate and execute optimized query plans, dynamically interacting with the execution engine to incorporate performance feedback for continuous improvement. 
Beyond optimizing query execution, CrackSQL~\cite{cracksql,zhou2025parrot} employs a local-to-global strategy for SQL dialect translation, iteratively guiding the LLM to fix failing sub-queries and progressively expanding to broader query segments until the entire query executes correctly.

\ptitle{Tool-Assisted Validation.}
Some systems further incorporate external tools, such as query rewrite engines or equivalence verifiers, to evaluate and validate optimized queries, providing a reward signal for iterative improvement. 
R-Bot~\cite{sun2025rbot} guides an LLM to select and sequence transformation rules for query rewriting, which are applied by an external engine and then evaluated for quality to guide refinement. 
Similarly, QUITE~\cite{song2025quite} uses an external verifier (\ie SQLSolver~\cite{sqlsolver}) to check query equivalence and defines a reward function based on execution metrics for progressive improvement.

\ptitle{Limitations.}
Despite leveraging environmental feedback or external verifiers to iteratively optimize query plans, these systems adhere to pipelines, transformation rules, and rewards meticulously designed by humans to ensure robust performance, incapable of autonomously exploring beyond procedural optimizations or adopting new optimization strategies.

\stitle{System Diagnosis.}
L2 data agents for system diagnosis are designed to actively perceive, localize, and resolve database performance issues through continuous monitoring and interaction, collecting evidence and reasoning over system states to generate actionable diagnoses.
Their employed feedback mechanisms are twofold, including direct execution feedback and further leveraging structured knowledge from graph-based repositories encoding prior diagnostic experiences.

\ptitle{Interactive Evidence Collection and Diagnosis.}
Interacting with database systems provides performance metrics (\eg estimated cost and storage impact) for reflection and iterative refinement.
DBdoctor~\cite{dbdoctor} monitors SQL execution and collects resource usage data to identify the root causes of performance issues, pinpoint problematic queries, and provide improvement suggestions.
More advanced agentic frameworks enhance this process:
D-Bot~\cite{zhou2024dbot} introduces a multi-agent system for database anomaly diagnosis, with a chief agent coordinating specialized agents, each aligned with a specific cluster of diagnostic knowledge extracted from documents. They collectively perform multi-step root cause analysis through a tree-search algorithm, continuously interacting with the database state to localize issues.
Similarly, Panda~\cite{singh2024panda} simulates database engineers' workflows, interacting with the database using four mechanisms: grounding to supply query context, verification with troubleshooting documents and tickets, affordance to estimate the performance impact, and feedback for continuous improvement.

\ptitle{Contextualized Diagnosis Knowledge.}
Some approaches further integrate evolving structured diagnosis knowledge graphs to accumulate experiences and memory over time. Data agents retrieve relevant subgraphs or identify missing connections to provide concrete, contextualized knowledge, enabling more informed diagnosis. 
DBAIOps~\cite{zhou2025dbaiops} builds a heterogeneous graph of diagnostic experience and employs an automatic graph evolution mechanism to explore relevant paths and detect potential gaps (\ie missing connections). This structured knowledge supports root cause identification and the generation of comprehensive diagnostic reports.

\ptitle{Limitations.}
While collecting evidence and localizing issues through database interactions and knowledge graphs significantly enhances performance and autonomy in anomaly diagnosis and resolution, these methods remain constrained by their reliance on structured diagnostic paths, human-defined worker agents, and predefined collaboration mechanisms. Consequently, they are unable to autonomously design diagnostic mechanisms to proactively address emerging anomalies and coordinate with tasks such as configuration tuning and query optimization, hindering the achievement of more efficient and robust data management.

\vspace{-0em}
\subsection{L2 Data Agents in Data Preparation}
\vspace{-0em}

\stitle{Data Cleaning.}
At L2, data agents can leverage their perception and interaction with environments to iteratively optimize cleaning strategies and output, therefore enabling partial autonomy. Relevant research can be distinguished by specific environments involved and feedback mechanisms.

\ptitle{Execution and Assessment Environments.}
One pattern at L2 involves treating code interpreters or data processing tools as environments for operational feedback. 
For instance, AutoPrep~\cite{fan2025autoprep} and CleanAgent~\cite{qi2025cleanagentautomatingdatastandardization} interact with execution engines (like Python or Docker) to receive feedback for refining generated code, with CleanAgent~\cite{qi2025cleanagentautomatingdatastandardization} further introducing a memory module to maintain historical conversation context.
AutoDCWorkflow~\cite{li2025autodcworkflow} executes cleaning operations on OpenRefine~\cite{openrefine} and subsequently uses the resulting internal data quality report as feedback for its iterative workflow.

\ptitle{Internal Self-Correction and Verification.}
Building on interaction with the execution environment, some L2 data agents further incorporate internal feedback loops for self-assessment and refinement. 
SketchFill~\cite{zhang2024sketchfill} and IterClean~\cite{IterClean} exemplify this. SketchFill~\cite{zhang2024sketchfill} includes an Evaluator and Reflector for its imputation formulas, while IterClean~\cite{IterClean} introduces a Repairer-Detector cycle where remaining errors trigger new repairs.
MegaTran~\cite{li2025megatran} interacts with a Python executor and uses its feedback to trigger a ``Sanity-check Reflection'' module or a Lazy-RAG mechanism to improve its generated transformation code iteratively.

\ptitle{Downstream Task Performance as Feedback.}
A more advanced form uses the performance of a downstream task as the ultimate environmental feedback. The framework proposed by Bendinelli et al.~\cite{bendinelli2025exploring} allows data agents to clean a dataset and submit it to a machine learning evaluation pipeline. The resulting model performance score serves as feedback, directly guiding subsequent cleaning actions to optimize the data for the specific downstream task.

\ptitle{Limitations.}
L2 data agents are still considered partially autonomous, as their actions are confined within human-designed workflows or pre-defined strategic pipelines. For instance, CleanAgent~\cite{qi2025cleanagentautomatingdatastandardization} and SketchFill~\cite{zhang2024sketchfill} execute and optimize within human-designed frameworks, not to devise pipelines. Moreover, their system architectures are closely coupled with specific data cleaning tasks (\eg CleanAgent's tailored column-type annotator~\cite {qi2025cleanagentautomatingdatastandardization}), resulting in limited capability to handle diverse and comprehensive tasks.

\stitle{Data Integration.}
Recent studies have introduced data agents with environmental perception and interaction to dynamically resolve ambiguities in resolution, optimize alignment strategies, and validate intermediate results during data integration. 

\ptitle{Execution-Driven Iterative Refinement.} 
SEED~\cite{chen2023seed} generates and validates code within sandboxed environments, reusing cached intermediate results for entity resolution. It also incorporates an optimizer that selectively deploys and configures LLM-assisted modules to reduce costs, and invokes data access tools based on observations on data retrieval outcomes to improve response accuracy.
COMEM~\cite{wang2025match} follows a ``filter–refine'' workflow where mid-sized LLMs screen candidates before stronger LLMs finalize matches, with positional-bias feedback dynamically refining the ranking process. 

\ptitle{Retrieval-Driven Iterative Refinement.} 
TaDA~\cite{huang2024transform} similarly employs tabular databases and vector indices to optimize table alignment strategies based on environmental feedback.
Agent-OM~\cite{qiang2024agent} improves by employing retrieval and matching agents that interact with hybrid databases and invoke alignment tools, enhanced by CoT planning and shared memory.
MILA~\cite{taboada2025ontology} integrates SBERT-based ontology retrieval with iterative LLM prompting, autonomously confirming high-confidence matches while adaptively handling ambiguous cases. 
Likewise, KG-RAG4SM~\cite{ma2025knowledge} and KCMF~\cite{xu2024kcmf} leverage external knowledge graphs to retrieve domain-specific subgraphs, refine semantic alignment, and mitigate hallucinations.

\ptitle{Limitations.}
Overall, L2 data agents represent a shift toward adaptive and self-improving data integration systems. However, their autonomy is limited within human-designed components, workflows, and task boundaries.

\stitle{Data Discovery.}
At L2, data agents actively engage with data systems to enable adaptive discovery in dynamic data lakes. Below, we present representative advances.

\ptitle{Direct Interaction with Data Repositories.}  
Some L2 agents directly interact with repositories via exploratory queries and schema traversal to dynamically uncover structural and semantic characteristics. 
LEDD~\cite{an2025ledd} actively queries large-scale data lakes via a federated access layer, querying metadata and analyzing content patterns in real-time to derive semantic representations, which are iteratively refined through a feedback loop that updates hierarchical catalogs and relation graphs.
DataVoyager~\cite{majumder2024data} interacts with live repositories by ingesting data, parsing schemas, and sampling content to perceive statistical patterns, which guide its autonomous analytical queries to explore latent structures.

\ptitle{Interaction via Execution Feedback.}  
Another approach emphasizes execution outcomes as feedback for refinement, where profiling quality or code execution results serve as environmental signals. 
MetaGen~\cite{alrubaye2025metadata} establishes validation feedback, such as hallucination detection and consistency checks, to reconstruct prompts and enhance metadata quality. 
DBDescGen~\cite{li2025autodb} proposes automatic database description generation for NL2SQL, in which execution results such as query success rate or accuracy serve as feedback to refine schema-level descriptions.
DATALORE~\cite{lou2024datalore} extends this feedback paradigm to data lineage reconstruction; it generates and executes transformation scripts, validates outputs, and iteratively adjusts logic based on success or mismatch signals.

\ptitle{Multimodal Data Discovery.}  
Studies also extend data discovery to heterogeneous and multimodal environments. 
Chorus~\cite{kayali2024chorus} utilizes foundation model embeddings to map schemas, documents, and ontological entities from structured, semi-structured, and unstructured sources into a shared representation space, enabling cross-source semantic alignment and bidirectional exchange of discovery signals across modalities. 
Wang et al.~\cite{wang2025towards} advance this by operationalizing cross-modal interaction via declarative query interfaces, multimodal operators, and data-driven query planning, enabling dispatch of sub-queries across modalities and aggregation of results.

\ptitle{Limitations.}
Despite these advances, L2 data agents remain bounded by human-defined workflows and coarse-grained feedback mechanisms. For instance, LEDD~\cite{an2025ledd} uses heuristic exploration without adaptive planning. Metadata and description generation frameworks~\cite{alrubaye2025metadata,li2025autodb} rely on coarse-grained success signals, resulting in unstable quality. 
These limitations highlight that L2 data agents operate within predefined frameworks, constrained in their ability to integrate fine-grained environmental feedback and lacking the capacity to autonomously orchestrate adaptive discovery pipelines.

\vspace{-0em}
\subsection{L2 Data Agents in Data Analysis}
\vspace{-0em}

\stitle{Structured Data Analysis.}
L2 data agents in structured data analysis enhance efficiency and insight extraction by autonomously interacting with tables or relational databases, executing code, and refining analyses through iterative feedback loops. They move beyond static querying to dynamically engage with, verify, and refine multi-step analytical processes, enabling more effective and complex data analysis, marking partial autonomy with limited strategic flexibility.

\etitle{TableQA}
In TableQA, data agents at L2 evolve from static responders to dynamic, adaptive problem-solvers.

\ptitle{Stateful and Adaptive TableQA.}
ReAcTable~\cite{ReAcTable2024} and Chain-of-Table~\cite{CHAINOFTABLE2024} leverage external execution for iterative reasoning, materializing the process into a series of observable state transitions where intermediate tables serve as feedback. ReAcTable~\cite{ReAcTable2024} employs a flexible ReAct~\cite{yao2023react}-inspired ``thought-action-observation'' cycle to generate and execute general-purpose code (\eg SQL or Python), using the observation to guide the next step. Chain-of-Table~\cite{CHAINOFTABLE2024} uses a structured ``plan-execute'' cycle with predefined atomic operations, where the transformed table itself is the stateful feedback.
In contrast, StructGPT~\cite{StructGPT2023} iteratively calls specialized APIs, using the returned data to extract evidence and plan further actions. TableMaster~\cite{cao2025tablemaster} combines semantic verbalization with an adaptive reasoning strategy that integrates textual and symbolic approaches to improve table understanding.
ST-Raptor~\cite{tang2026straptor} further proposes a tree-based QA framework for semi-structured tables, constructing a Hierarchical Orthogonal Tree to encode structure and using a two-stage validation strategy for robust answers.

\ptitle{Multi-Agent TableQA.}
Multi-agent collaboration is also used to enhance TableQA. AutoTQA~\cite{AutoTQA2024} delineates roles such as Planner, Engineer, and Critic to manage planning, execution, evaluation, and revision, enabling strategic replanning. 
Likewise, Table-Critic~\cite{TableCritic2025} employs a cooperative framework including Judge, Critic, and Refiner modules for multi-round error correction. It uniquely features a self-evolving template tree managed by a Curator, allowing the system to learn from experience by distilling correction patterns into reusable critique templates for continuous refinement.

\ptitle{Limitations.}
These studies represent a more advanced form of adaptive adjustment tailored for TableQA. However, these data agents at L2 exhibit several limitations. Their feedback is often superficial (\eg raw outputs or errors) rather than reflective, leading to tactical, symptomatic refinements that can result in unproductive loops without addressing root-cause strategic errors~\cite{AdaPlanner2023, TableCritic2025}. Furthermore, they depend heavily on predefined tools and APIs, limiting their ability to handle unforeseen problems~\cite{masterman2024landscapeemergingaiagent}. For example, Table-Critic~\cite{TableCritic2025} operates within rigid, human-designed architectures, restricting flexibility and versatility.

\etitle{NL2SQL}
L2 data agents exhibit partial autonomy in NL2SQL, characterized by their ability to perceive and interact with databases~\cite{deng2025reforce, zhu2026sqlconductor}, leverage SQL execution feedback~\cite{macsql, zhu2025elliesql}, and decompose complex SQL reasoning~\cite{pourreza2025chasesql, li2025alphasql}.

\ptitle{Execution Feedback-Driven SQL Refinement.}
Earlier efforts have explored utilizing execution feedback. MAC-SQL~\cite{macsql} introduces a multi-agent framework with a refiner using SQL execution feedback to iteratively refine faulty queries.
OpenSearch-SQL~\cite{opensearchsql} employs a consistency alignment module to reduce instruction-following failures and hallucinations. 
ReFoRCE~\cite{deng2025reforce} extends this by actively interacting with databases through iterative column exploration guided by execution feedback, enabling progressive self-improvement.

\ptitle{Planning and Decomposition.}
Some studies further incorporate meticulously designed reasoning pipelines for planning and decomposition to improve performance on complex queries. 
Chase-SQL~\cite{pourreza2025chasesql} proposes a query plan CoT and a divide-and-conquer strategy to decompose SQL generation into subtasks, subsequently refined via database interaction.
ChatBI~\cite{lian2024chatbi} decomposes schema linking into a view selection problem to handle large schemas.
DeepEye-SQL~\cite{li2025deepeyesql} implements an orchestrated pipeline inspired by the software development life cycle, decomposing SQL generation and deploying a toolchain of checkers with interactive feedback.

\ptitle{Improvement and Limitations.}
More advanced studies have moved toward more flexible approaches. For example, Alpha-SQL~\cite{li2025alphasql} employs Monte Carlo Tree Search to explore effective action combinations during NL2SQL translation. While this reduces reliance on human orchestration, it essentially depends on search heuristics rather than deliberate pipeline orchestration. Furthermore, its manually designed, task-specific actions tailored for NL2SQL limit applicability to a broader range of data-related tasks, bounding its autonomy.

\etitle{NL2VIS}
To better manage NL2VIS tasks, data agents at L2 can execute code, assess the visual output, and use the feedback to debug, self-correct, and improve the final visualization within structured, carefully orchestrated pipelines, such as multi-agent systems or iterative refinement cycles, to handle the complexities of visualization generation.

\ptitle{Iterative and Collaborative Refinement.}
For example, MatPlotAgent~\cite{MatPlotAgent2024ACL} utilizes iterative refinement with a dual-feedback mechanism: iterative code debugging and a visual feedback loop incorporating a multimodal ``visual agent'' that inspects charts to suggest corrections for non-code errors. 
nvAgent~\cite{nvAgent2025ACL} also employs a multi-agent ``divide-and-conquer'' mechanism to handle complex queries, using specialized agents to interact with the database to filter schemas, plan the visualization step-by-step, and validate code.

\ptitle{Self-Improvement.}
Some systems leverage automated mechanisms for continuous learning. Text2Chart31~\cite{Text2Chart312024EMNLP} introduces reinforcement learning-based instruction tuning with automatically generated reward signals, enabling scalable, closed-loop improvement without constant human feedback. It also presents a dataset with a wider range of chart types, covering 3D and volumetric plots, to train more versatile agents.

\ptitle{Explainability.}
DeepVIS~\cite{shuai2025deepvis} integrates CoT reasoning into the NL2VIS pipeline and features an interactive visual interface, allowing users to inspect the step-by-step reasoning and provide targeted feedback.

\ptitle{Limitations.}
Despite these advances, these systems primarily operate within predefined procedural workflows tailored for NL2VIS tasks. They excel at refining and validating visualizations within a defined scope but lack the autonomous, high-level reasoning to interpret complex user goals and design comprehensive strategies that span the full data lifecycle.

\stitle{Unstructured Data Analysis.}
In unstructured data analysis, L2 data agents evolve from passive processing to active, strategic information handling. They dynamically interact with complex documents, addressing core challenges like extreme length, factual grounding, multimodality, and high analysis cost through intelligent navigation~\cite{crag2024}, dynamic information seeking, and resource-aware optimization.

\etitle{Textual Documents.}
A primary challenge is managing vast, noisy textual documents that exceed typical context windows. 

\ptitle{Structured Representation for Navigation.}
Some studies replace linear processing with intelligent navigation. For example, WebGPT~\cite{nakano2022webgpt} demonstrates training LLMs to operate a browsing environment, autonomously search, select, and cite sources to produce more faithful long-form answers. Subsequent systems create structured internal representations: ReadAgent~\cite{readagent2024} uses a gist-based memory for text compression and selective inspection, while GraphReader~\cite{graphreader2024} models documents as graphs for targeted traversal by a planning agent.

\ptitle{Uncertainty-Driven Verification.}
Beyond navigation, other approaches introduce dynamic, self-driven verification. Systems like Self-RAG~\cite{asai2023selfrag} and FLARE~\cite{flare2023} assess uncertainty, allowing them to pause, trigger retrievals of additional information, and critique their own outputs. REAR~\cite{rear2024} further ensures robustness by incorporating explicit assessment modules to evaluate the relevance of retrieved documents.

\etitle{Multimodal Documents.}
L2 agents also advance the analysis of multimodal documents integrating text, visual layouts, tables, and images. 
MACT~\cite{yu2025mact} uses a multi-agent framework with a specialized judgment agent for verification and revision. 
Complementing this, DataPuzzle~\cite{zhang2025datapuzzle} proposes a structure-first approach, transforming unstructured multimodal inputs into interpretable representations (\eg tables or graphs). This enables exploration, extraction, and reasoning over structured data, overseen by a meta-agent that coordinates the process and triggers revisions, enhancing transparency and verifiability.

\etitle{DocumentDB.}
To address the high cost and latency of LLM-based extraction, database-inspired systems (DocumentDB) are developed to treat unstructured data analysis as a query optimization problem.
QUEST~\cite{sun2025quest} introduces an ``optimize at execution time'' architecture to create tailored query plans at execution time, ordering filters based on cost and selectivity. 
Similarly, Doctopus~\cite{chai2025doctopus} provides a budget-aware framework that dynamically selects the most cost-effective extraction strategy based on user constraints and quality estimates.

\ptitle{Limitations.}
Collectively, L2 data agents for unstructured data analysis are more strategic, robust, and efficient, managing longer and more complex data, verifying information, and optimizing for cost. However, their autonomy remains limited as they typically rely on predefined tools, operators, and manually curated data pipelines, which sets the stage for future advancements toward more comprehensive autonomy.

\vspace{-.25em}
\stitle{Report Generation.}
In report and data narrative generation, L2 data agents complete a ``generate-visualize-detect-refine'' cycle through perceiving and interacting with databases, tools, visual interfaces, etc.

\ptitle{Narrative Generation and Verification.}
One research direction focuses on autonomous report generation and data storytelling via multimodal agentic architectures. 
Multimodal DeepResearcher~\cite{yang2025deepresearcher} adopts a four-stage pipeline: researching, exemplar textualization, planning, and multimodal synthesis, grounded in a Formal Description of Visualization (FDV) to produce interleaved text–chart reports. 
DAgent~\cite{xu2025dagent} decomposes relational database queries to synthesize reports from execution results, while Data Director~\cite{shen2024datadirector} extends storytelling to data video creation. More recently, DataMagic~\cite{xie2026datamagic,xie2026datamagicvldb} transforms tabular data into narrated data videos through a declarative specification (DVSpec) that decouples scene design from rendering, and a ``generate-then-orchestrate'' multi-agent pipeline that synchronizes narration with animation.
To improve factual accuracy, systems emphasize verification. DataNarrative~\cite{islam2024datanarrative} employs a generator-evaluator pair to iteratively validate chart-text narratives, and ChartInsighter~\cite{wang2025chartinsighter} combines multi-agent tool use with self-consistency to reduce hallucination.

\ptitle{Interactive Visual Analytics.}
Another direction is to reshape user interaction with visualization systems.
LightVA~\cite{zhao2025lightva} assists users with a lightweight ``planner-executor-controller'' loop for task decomposition, execution, and refinement. 
ProactiveVA~\cite{zhao2025proactiveva} employs a UI agent to monitor interactions and proactively offer context-aware analytic suggestions.
Others act as mediators: NLI4VolVis~\cite{ai2025nli4volvis} enables conversational exploration of volumetric data, and VOICE~\cite{jia2024voice} coordinates ``pack of bots'' to interpret queries, assign subtasks, and generate synchronized visual explanations.

\ptitle{Limitations.}
Collectively, these efforts advance beyond the ``prompt-response'' paradigm, achieving partial autonomy through ``generate-visualize-detect-refine'' cycles. However, their autonomy remains bounded by predefined pipelines and agentic modules, narrow task scopes, and a lack of comprehensive orchestration capabilities.

\vspace{-.0em}
\subsection{The Glass Ceiling of L2 Data Agents}
\vspace{-.0em}
L2 data agents integrate environmental perception with autonomous procedural execution to achieve partial autonomy in data-related tasks. However, as discussed, they face a glass ceiling marked by critical limitations as follows.

\ptitle{Dependence on human-defined pipelines.} 
L2 data agents execute and optimize outputs within pre-established pipelines but lack the ability to independently orchestrate pipelines tailored to high-level or evolving task goals. 
For example, CleanAgent~\cite{qi2025cleanagentautomatingdatastandardization} follows a human-crafted architecture with four specified components and pipelines for data cleaning rather than autonomously orchestrating the entire process.

\ptitle{Task-specific rigidity.}
At L2, data agents are closely tied to specific domains, restricting generalizability. 
For instance, Chase-SQL~\cite{pourreza2025chasesql} designs a pipeline specialized for NL2SQL translation and query-fixing and fine-tunes an LLM solely for SQL candidate selection. This approach can not be adopted for a broader range of data-related tasks, such as data integration or more comprehensive tasks spanning data management, preparation, and analysis. Such narrow task specialization limits their effectiveness in addressing the diverse and comprehensive challenges common in real-world data-related tasks.

Together, the constraints of pipeline dependence and task specificity reveal the autonomy boundary of L2 data agents. As summarized in Table~\ref{tab:l2_comparison}, L2 systems widely exploit environmental perception, tool invocation, and feedback-driven refinement, but these capabilities are mostly exercised within task-specific and human-designed pipelines. Therefore, L2 agents advance from static responders to procedural executors, yet still fall short of self-directed orchestration over comprehensive data-related tasks.

\section{L3: Striving for Autonomous Data Agent}
\label{sec:l3}

\begin{figure}[t!]
    \centering
    \includegraphics[width=\linewidth]{\paperroot 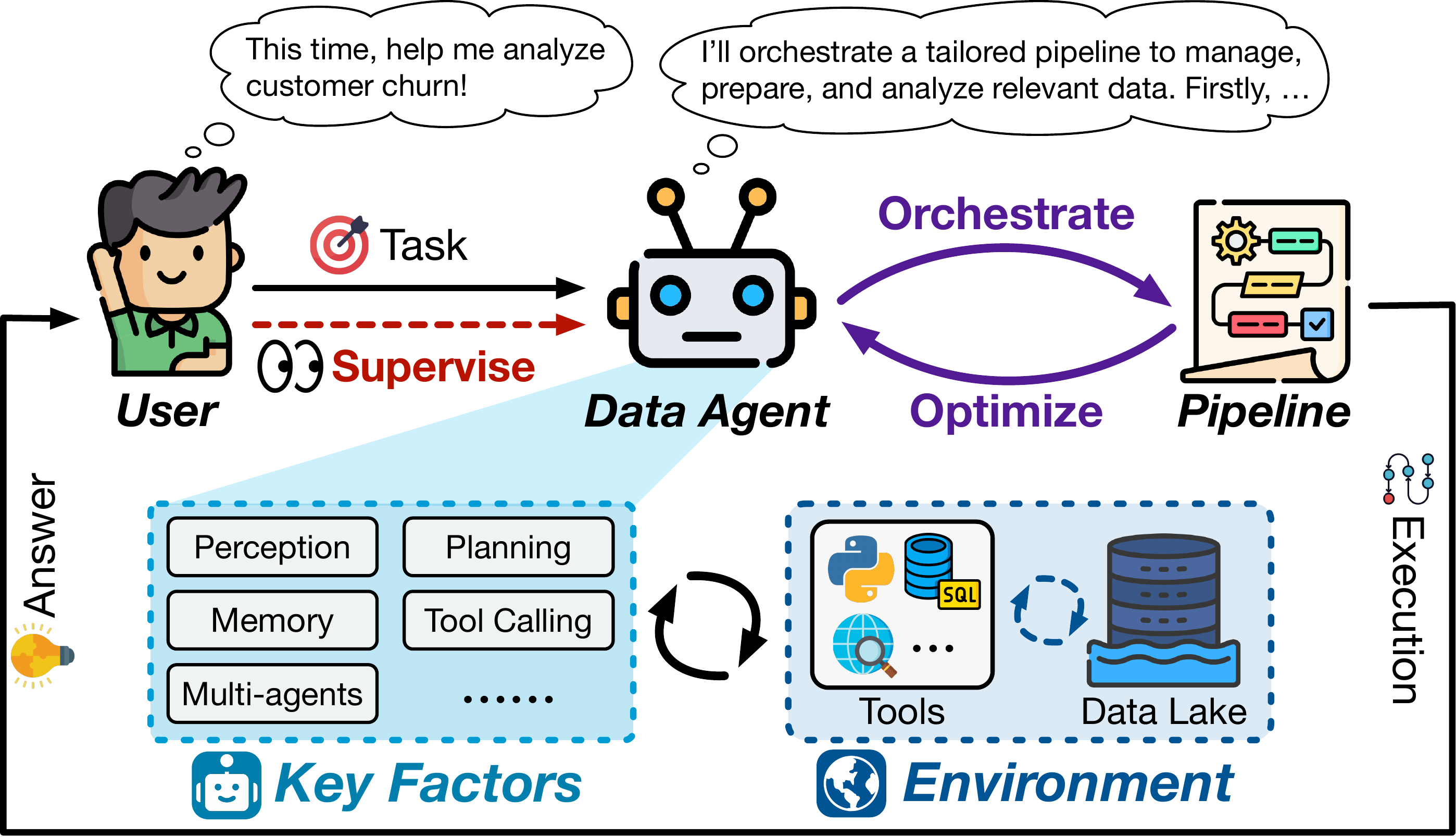}
    \vspace{-0em}
    \caption{ L3 Data Agents (Conditional Autonomy).}
    \vspace{-0em}
    \label{fig:l3_agent}
\end{figure}

\begin{table*}[t!]
\centering
\caption{Comparison of Representative Proto-L3 Data Agents from Academia Research and Industry Products. Compares Open-source: availability; Undef Ops.: capabilities in utilizing unpredefined operators; data-related task coverage across data management, preparation, analysis; data complexity dimensions: Multi-source (Multis.), Heterogeneous (Hete.), and Multimodal (Multim.). ${^\dagger}$ indicates industry system.}
\vspace{-0.0em}
\label{tab:l3_comparison}
\resizebox{\textwidth}{!}{\begin{tabular}{|c|c|c|ccc|ccc|ccc|ccc|}
\hline
\multirow{3}{*}{\textbf{Data Agent}} &
  \multirow{3}{*}{\textbf{\begin{tabular}[c]{@{}c@{}}Open-\\ source\end{tabular}}} &
  \multirow{3}{*}{\textbf{\begin{tabular}[c]{@{}c@{}}Undef\\ Ops.\end{tabular}}} &
  \multicolumn{3}{c|}{\textbf{Data Complexity}} &
  \multicolumn{3}{c|}{\textbf{Data Management}} &
  \multicolumn{3}{c|}{\textbf{Data Preparation}} &
  \multicolumn{3}{c|}{\textbf{Data Analysis}} \\ \cline{4-15} 
 &
   &
   &
  \multicolumn{1}{c|}{Multis.} &
  \multicolumn{1}{c|}{Hete.} &
  Multim. &
  \multicolumn{1}{c|}{\begin{tabular}[c]{@{}c@{}}Config \\[-1.5pt] Tun.\end{tabular}} &
  \multicolumn{1}{c|}{\begin{tabular}[c]{@{}c@{}}Query \\[-1.5pt] Opt.\end{tabular}} &
  \begin{tabular}[c]{@{}c@{}}Sys. \\[-1.5pt] Diag.\end{tabular} &
  \multicolumn{1}{c|}{\begin{tabular}[c]{@{}c@{}}Data \\[-1.5pt] Clean.\end{tabular}} &
  \multicolumn{1}{c|}{\begin{tabular}[c]{@{}c@{}}Data \\[-1.5pt] Integ.\end{tabular}} &
  \begin{tabular}[c]{@{}c@{}}Data \\[-1.5pt] Disc.\end{tabular} &
  \multicolumn{1}{c|}{Struct.} &
  \multicolumn{1}{c|}{Unstruct.} &
  \begin{tabular}[c]{@{}c@{}}Report \\[-1.5pt] Gen.\end{tabular} \\ \hline
DeepEye~\cite{li2026deepeyedemo} &
  \greencheck &
  - &
  \multicolumn{1}{c|}{\greencheck} &
  \multicolumn{1}{c|}{\greencheck} &
  \greencheck &
  \multicolumn{1}{c|}{-} &
  \multicolumn{1}{c|}{-} &
  - &
  \multicolumn{1}{c|}{\greencheck} &
  \multicolumn{1}{c|}{\greencheck} &
  \greencheck &
  \multicolumn{1}{c|}{\greencheck} &
  \multicolumn{1}{c|}{\greencheck} &
  \greencheck \\ \hline
AgenticData~\cite{agenticdata} &
  - &
  \halfcheck &
  \multicolumn{1}{c|}{\greencheck} &
  \multicolumn{1}{c|}{\greencheck} &
  - &
  \multicolumn{1}{c|}{-} &
  \multicolumn{1}{c|}{\greencheck} &
  \greencheck &
  \multicolumn{1}{c|}{\greencheck} &
  \multicolumn{1}{c|}{\greencheck} &
  \greencheck &
  \multicolumn{1}{c|}{\greencheck} &
  \multicolumn{1}{c|}{\greencheck} &
  - \\ \hline
DeepAnalyze~\cite{deepanalyze} &
  \greencheck &
  - &
  \multicolumn{1}{c|}{\greencheck} &
  \multicolumn{1}{c|}{\greencheck} &
  - &
  \multicolumn{1}{c|}{-} &
  \multicolumn{1}{c|}{-} &
  - &
  \multicolumn{1}{c|}{\greencheck} &
  \multicolumn{1}{c|}{\greencheck} &
  \greencheck &
  \multicolumn{1}{c|}{\greencheck} &
  \multicolumn{1}{c|}{\greencheck} &
  \greencheck \\ \hline
AOP~\cite{wang2025aop} &
  - &
  - &
  \multicolumn{1}{c|}{\greencheck} &
  \multicolumn{1}{c|}{\greencheck} &
  \greencheck &
  \multicolumn{1}{c|}{-} &
  \multicolumn{1}{c|}{\greencheck} &
  \greencheck &
  \multicolumn{1}{c|}{\greencheck} &
  \multicolumn{1}{c|}{\greencheck} &
  \greencheck &
  \multicolumn{1}{c|}{\greencheck} &
  \multicolumn{1}{c|}{\greencheck} &
  - \\ \hline
iDataLake~\cite{wang2025idatalake} &
  - &
  - &
  \multicolumn{1}{c|}{\greencheck} &
  \multicolumn{1}{c|}{\greencheck} &
  \greencheck &
  \multicolumn{1}{c|}{-} &
  \multicolumn{1}{c|}{\greencheck} &
  - &
  \multicolumn{1}{c|}{\greencheck} &
  \multicolumn{1}{c|}{\greencheck} &
  \greencheck &
  \multicolumn{1}{c|}{\greencheck} &
  \multicolumn{1}{c|}{\greencheck} &
  \greencheck \\ \hline
Data Interpreter~\cite{data-interpreter} &
  \greencheck &
  - &
  \multicolumn{1}{c|}{-} &
  \multicolumn{1}{c|}{\greencheck} &
  \greencheck &
  \multicolumn{1}{c|}{-} &
  \multicolumn{1}{c|}{-} &
  - &
  \multicolumn{1}{c|}{\greencheck} &
  \multicolumn{1}{c|}{-} &
  \greencheck &
  \multicolumn{1}{c|}{\greencheck} &
  \multicolumn{1}{c|}{\greencheck} &
  \greencheck \\ \hline \hline
 \begin{tabular}[c]{@{}c@{}}${^\dagger}$JoyAgent~\cite{JoyAgent-JDGenie}\end{tabular} &
  \halfcheck &
  \halfcheck &
  \multicolumn{1}{c|}{\greencheck} &
  \multicolumn{1}{c|}{\greencheck} &
  - &
  \multicolumn{1}{c|}{-} &
  \multicolumn{1}{c|}{-} &
  \greencheck &
  \multicolumn{1}{c|}{\greencheck} &
  \multicolumn{1}{c|}{\greencheck} &
  \greencheck &
  \multicolumn{1}{c|}{\greencheck} &
  \multicolumn{1}{c|}{\greencheck} &
  \greencheck \\ \hline
\begin{tabular}[c]{@{}c@{}}${^\dagger}$BigQuery~\cite{google_bigquery}\end{tabular} &
  - &
  - &
  \multicolumn{1}{c|}{\greencheck} &
  \multicolumn{1}{c|}{\greencheck} &
  - &
  \multicolumn{1}{c|}{-} &
  \multicolumn{1}{c|}{\greencheck} &
  - &
  \multicolumn{1}{c|}{\greencheck} &
  \multicolumn{1}{c|}{\greencheck} &
  \greencheck &
  \multicolumn{1}{c|}{\greencheck} &
  \multicolumn{1}{c|}{-} &
  - \\ \hline
\begin{tabular}[c]{@{}c@{}}${^\dagger}$Cortex~\cite{snowflake_cortex}\end{tabular} &
  - &
  - &
  \multicolumn{1}{c|}{\greencheck} &
  \multicolumn{1}{c|}{\greencheck} &
  \greencheck &
  \multicolumn{1}{c|}{-} &
  \multicolumn{1}{c|}{-} &
  - &
  \multicolumn{1}{c|}{\greencheck} &
  \multicolumn{1}{c|}{\greencheck} &
  \greencheck &
  \multicolumn{1}{c|}{\greencheck} &
  \multicolumn{1}{c|}{\greencheck} &
  - \\ \hline
\begin{tabular}[c]{@{}c@{}}${^\dagger}$Xata Agent~\cite{xata_agent}\end{tabular} &
  - &
  - &
  \multicolumn{1}{c|}{\greencheck} &
  \multicolumn{1}{c|}{\greencheck} &
  - &
  \multicolumn{1}{c|}{\greencheck} &
  \multicolumn{1}{c|}{\greencheck} &
  \greencheck &
  \multicolumn{1}{c|}{-} &
  \multicolumn{1}{c|}{-} &
  \greencheck &
  \multicolumn{1}{c|}{-} &
  \multicolumn{1}{c|}{-} &
  - \\ \hline
 \begin{tabular}[c]{@{}c@{}}${^\dagger}$SiriusBI~\cite{SiriusBI}\end{tabular} &
  - &
  - &
  \multicolumn{1}{c|}{-} &
  \multicolumn{1}{c|}{-} &
  - &
  \multicolumn{1}{c|}{-} &
  \multicolumn{1}{c|}{-} &
  - &
  \multicolumn{1}{c|}{\greencheck} &
  \multicolumn{1}{c|}{-} &
  \greencheck &
  \multicolumn{1}{c|}{\greencheck} &
  \multicolumn{1}{c|}{-} &
  \greencheck \\ \hline
\end{tabular}}
\vspace{-0em}
\end{table*}

\subsection{The Critical Leap: From ``Executor'' to ``Dominator''}
In this section, we elaborate on the transformative advancement of data agents to the L3, where the leap to L3 represents a critical transfer of dominance in data-related tasks, achieving conditional autonomy under supervision and signaling the emergence of true autonomy. 

As illustrated in Figure~\ref{fig:l3_agent}, L3 data agents are expected to autonomously orchestrate and optimize pipelines rather than following human-defined ones; managing diverse and comprehensive tasks that potentially span the entire data lifecycle from data management and preparation to analysis, rather than isolated and task-specific procedures.
Formally, at L3, data agent $\mathcal{A}$ autonomously manages the entire pipeline from orchestration $\pi_{\mathcal{A}}$ to execution $\epsilon_{\mathcal{A}}$, tackling versatile and comprehensive data-related tasks $\mathcal{T}$ under human $\mathcal{H}$ supervision:
\vspace{-0em}
\begin{equation*}
\begin{aligned}
\mathcal{A} &: \pi_{\mathcal{A}}(\mathcal{T}, \mathcal{D}, \mathcal{E}, \mathcal{M}) \rightarrow P; ~~\epsilon_{\mathcal{A}}(P, \mathcal{D}, \mathcal{E}, \mathcal{M}) \rightarrow \mathcal{O} \\
\mathcal{H} &: \text{Supervise}(\pi_{\mathcal{A}}, \epsilon_{\mathcal{A}})
\end{aligned}
\end{equation*}
\vspace{-0em}

Given a high-level goal such as ``analyze customer churn'', L3 data agents are expected to orchestrate and coordinate an effective pipeline from data management and preparation to analysis to achieve it. However, this autonomy remains conditional, as humans retain a supervisory role.

Developing L3 data agents remains an ongoing research frontier. To date, no existing system has fully realized such versatile, self-directed orchestration capabilities that define a complete L3 data agent. However, emerging efforts from both academia and industry
are beginning to address these challenges, giving rise to what we term ``Proto-L3'' data agents.
While they do not yet achieve the full spectrum of L3, they exhibit exploratory steps and promising potential toward this goal. 
The following section analyzes these pioneering efforts, which are summarized in Table~\ref{tab:l3_comparison}.

\vspace{-0em}
\subsection{Emerging Efforts and Proto-L3 Systems}
\vspace{-0em}
Emerging efforts on Proto-L3 data agents aim to transform manually designed components and processing procedures into autonomously orchestrated and optimized pipelines for diverse and complex data-related tasks, as illustrated in Figure~\ref{fig:l3_agent}.

\ptitle{Orchestrating Pipeline.}
Data Interpreter~\cite{data-interpreter} recasts orchestration as Hierarchical Graph Modeling. It autonomously decomposes high-level tasks into a structured Task Graph and an executable Action Graph, which is dynamically modified via Iterative Graph Refinement based on graph executor feedback.

\ptitle{Broadening Task Scope.}
However, these systems are limited in task scope, primarily concentrating on data analysis and relying heavily on preprocessed data. 
SiriusBI~\cite{SiriusBI} incorporates Data Insight and Knowledge Management modules, automating basic data preparation and metadata management for orchestration of multi-agent workflows for complex analytical queries.
iDataLake~\cite{wang2025idatalake} and AOP~\cite{wang2025aop} further extend to heterogeneous data lakes. 
iDataLake~\cite{wang2025idatalake} orchestrates pipelines from semantic operators and uses a unified embedding space for multimodal data linking. 
Similarly, AOP~\cite{wang2025aop} utilizes predefined semantic operators for data preparation and semantic analytics across heterogeneous data, featuring a pipeline optimizer that enables real-time, reflective adjustments to enhance efficiency. DeepAnalyze~\cite{deepanalyze} introduces a curriculum-based agentic reinforcement learning to progressively train data agents to adaptively orchestrate data preparation and analysis through five predefined actions. DeepEye~\cite{li2026deepeyedemo} further adopts a workflow-centric architecture with unified multimodal orchestration to integrate structured and unstructured sources, and employs a database-inspired workflow engine to compile, validate, optimize, and execute the orchestrated analytical workflows.

\ptitle{Industrial Practice.}
Recently, this field has also attracted significant attention from the industry. Snowflake Cortex~\cite{snowflake_cortex} orchestrates over structured and unstructured data sources through a specialized Cortex search service to deliver insights. 
Google's BigQuery~\cite{google_bigquery} focuses on structured data, providing query optimization to enhance performance and facilitate data preparation and analysis.
Xata Agent~\cite{xata_agent} offers comprehensive data management by orchestrating configuration tuning, query optimization, and system diagnosis to support data discovery.

\ptitle{Overcoming Predefined Toolsets and Operators.}
Despite advances in pipeline orchestration, a key limitation is the reliance on pre-defined agentic components~\cite{bytedance_data_agents, databricks_dsa}, collaboration mechanisms~\cite{SiriusBI}, or tools and operators~\cite{data-interpreter, google_bigquery, snowflake_cortex}. 
JoyAgent~\cite{JoyAgent-JDGenie} begins to overcome this via ``Tool Evolution'', dynamically creating new tools by recombining atomic ones, and uses a multi-level thinking framework to decompose queries into executable Directed Acyclic Graphs.
AgenticData~\cite{agenticdata} supports both predefined operators and non-predefined ones curated via LLM-based code generation. It broadens task scope to data management, preparation, and analysis across heterogeneous sources via customized Model-Context Protocol (MCP) servers and employs a feedback-driven, multi-agent planner to create tree-structured semantic pipelines. 
\ptitle{Evaluation in Complex Data Environments.}
As state-of-the-art systems move toward autonomous orchestration in complex and diverse data environments, recent benchmarks have begun to examine more comprehensive and challenging data-agent capabilities across data discovery, preparation, coding, reasoning, and analysis.
DABstep~\cite{Egg2025DABstep} evaluates realistic multi-step data analysis with code-based processing and contextual reasoning, while CoDA-Bench~\cite{zhang2026codabench} stresses data discovery, manipulation, and analysis in noisy sandboxed environments.
FDABench~\cite{wang2026fdabench} targets analytical queries over heterogeneous data sources, and DAComp~\cite{lei2026dacomp} covers enterprise data-intelligence workflows spanning data engineering and open-ended analysis.
More recently, AgenticDataBench~\cite{sun2026agenticdatabench} provides fine-grained skill-level diagnosis across diverse data-science tasks.
These benchmarks make evaluation closer to real-world data-agent settings, where success depends not only on final answer accuracy but also on discovering relevant data, coordinating tools, grounding intermediate steps, and sustaining reliable workflows with reduced human intervention.

\ptitle{Limitations.}
Collectively, these developments signal a significant shift toward agent-led orchestration in data-related tasks. However, they have not yet achieved true L3 autonomy, since task dominance has not been reliably transferred to the agent.
Currently, these Proto-L3 systems typically depend on predefined operators, exhibit constrained task scope (\eg limited data management capabilities), focus on tactical rather than strategic reasoning, and lack genuine self-evolution.
Their orchestration is also still bounded by action spaces, validation signals, and heuristic failure-handling mechanisms, making it difficult to reliably justify and take responsibility for errors in complex data environments.
Therefore, accountability remains externally anchored in the human/system design process rather than being assumed by the agent itself.
Bridging these gaps represents a critical challenge for realizing L3 data agents.

\vspace{-0em}
\subsection{Challenges and Research Opportunities Towards True L3}
\vspace{-0em}

Despite promising Proto-L3 advancements, significant gaps remain toward true L3 autonomy, highlighting fundamental challenges and research opportunities.

\ptitle{1) Limited Autonomy in Pipeline Orchestration.}
A primary limitation of current Proto-L3 data agents is reliance on predefined operations or tools. For instance, AOP~\cite{wang2025aop}, BigQuery~\cite{google_bigquery}, and Cortex~\cite{snowflake_cortex} orchestrate pipelines with predefined semantic operators, which restricts their autonomy. While JoyAgent~\cite{JoyAgent-JDGenie} features ``Tool Evolution'' to reassemble tools, such recombination is still limited within the atomic tool pool. 
AgenticData~\cite{agenticdata} generates non-predefined operations via LLM-based code generation, but lacks validation regarding the effectiveness and robustness. 
Sun et al.~\cite{sun2025data} discuss the feasibility of automatic data-skill discovery to create a dynamic toolkit beyond fixed operator sets. Future work should integrate continuous skill discovery with orchestration to generate, evaluate, and deploy emergent skills, further advancing autonomy toward true L3.

\ptitle{2) Incomplete Coverage of the Data Lifecycle.}
Current Proto-L3 systems are significantly constrained by a relatively narrow scope in data-related tasks, primarily centered on data analysis~\cite{data-interpreter} or very limited data preparation~\cite{SiriusBI, JoyAgent-JDGenie, wang2025aop}. AgenticData~\cite{agenticdata} attempts to support management across heterogeneous sources through customized MCP servers, but its capabilities remain confined to basic monitoring and manipulation.
However, a true L3 data agent must be a versatile ``data expert'' spanning the entire lifecycle. Existing systems largely neglect crucial data management tasks such as configuration tuning, query optimization, and system diagnosis, limiting them to downstream analytical processes on ready-to-use data. The key research opportunity is enhancing versatility to cover the full data lifecycle, from system management to data preparation and insight extraction.

\ptitle{3) Deficiencies in Advanced Reasoning.}
Current Proto-L3 data agents exhibit strong tactical capabilities, effectively addressing immediate errors and refining recent steps. However, they lack advanced reasoning to reassess and adapt their overarching analysis strategy. 
By focusing narrowly on the symptom of a failed step, current systems are easily trapped in unproductive loops rather than diagnosing and resolving root causes.
Advancing requires research into causal reasoning, meta-reasoning for cross-process optimization, and sophisticated memory architectures for abstract strategic knowledge.

\ptitle{4) Inadequate Adaptation to Dynamic Environment.}
Real-world data ecosystems are dynamic, shaped by data drift and shifting schemas, yet current systems are operated and evaluated against static data without self-evolution as environments evolve.
While most existing works rely on inherent reasoning, DeepAnalyze~\cite{deepanalyze} employs agentic reinforcement learning to enhance adaptivity, though requiring substantial human effort. Therefore, developing more effective and human-free methods to enable data agents' adaptation to evolving environments, together with robust evaluation under dynamic conditions, represents an important research direction.

\section{L4--L5: Vision of Proactive and Generative Data Agent}
\label{sec:l4_l5}

While current research explores the L2-to-L3 transition, L4 and L5 represent a vision for data agents, marking a profound shift from conditional, task-driven autonomy to proactive self-governance and ultimately generative innovation. 
Looking forward, it is crucial to envision not only the capabilities these agents are expected to achieve but also essential advancements that enable their evolution.

\vspace{-0em}
\subsection{L4: High Automation}
\vspace{-.0em}

\begin{figure}[t!]
    \centering
    \includegraphics[width=\linewidth]{\paperroot 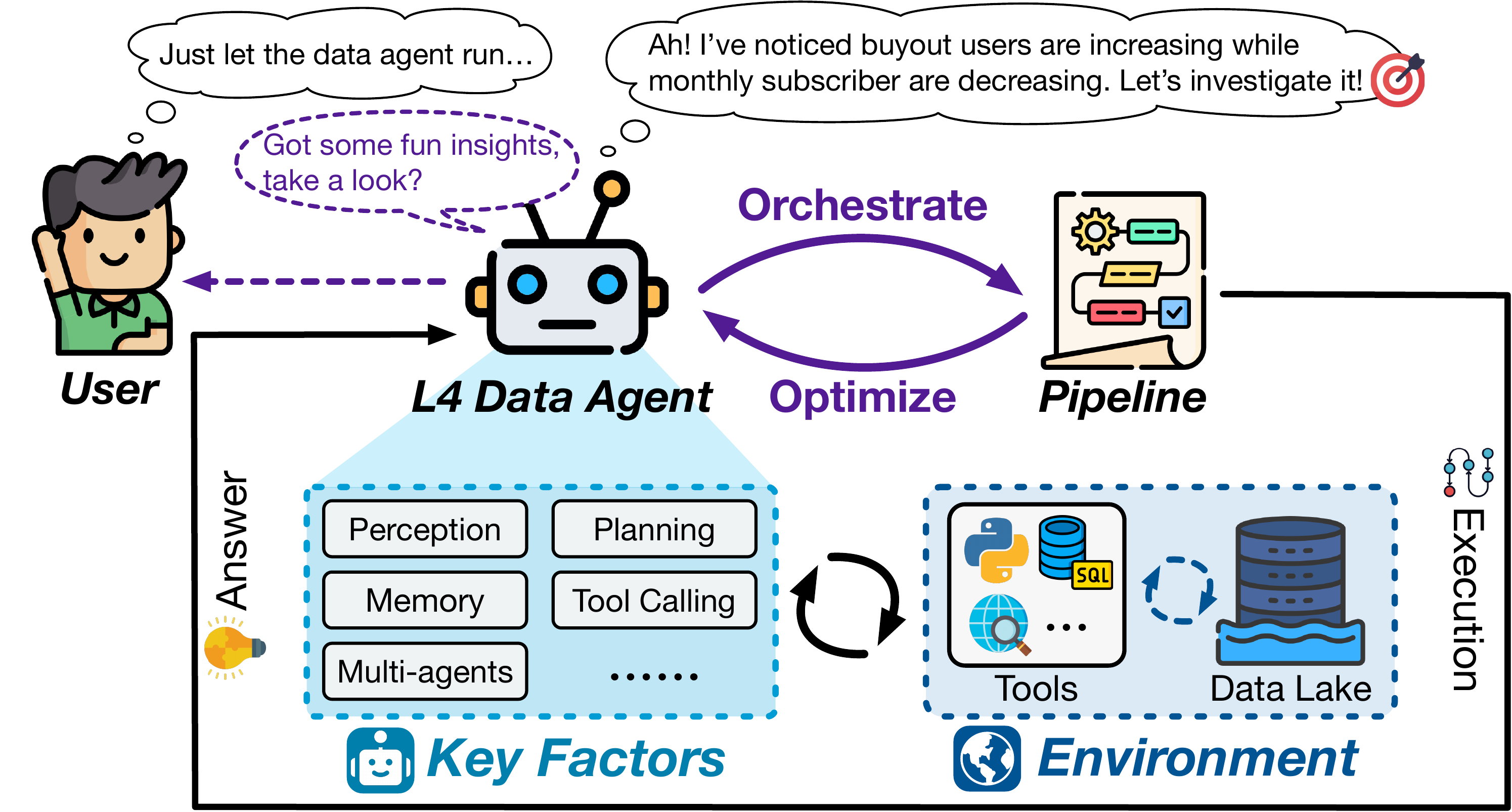}
    \vspace{-0em}
    \caption{L4 Data Agents (High Autonomy).}
    \vspace{-0em}
    \label{fig:l4_agent}
\end{figure}

At L4, data agents are envisioned to achieve a high degree of autonomy, being fully delegated the responsibility for data-related tasks. 
Operating independently of human supervision or explicit instructions, they autonomously monitor and explore data lakes to proactively identify valuable and emerging tasks, rather than simply responding to given goals,
shifting the user's role from active supervisor to passive onlooker and recipient of insights, as shown by Figure~\ref{fig:l4_agent}. 
For instance, it can detect an emerging trend in user behavior, like the increase in buyout subscriptions and the decrease in monthly subscriptions, then initiate a root-cause analysis for data drift, or re-optimize an indexing strategy as needed.

Formally, the data agent $\mathcal{A}$ takes full initiative, not only orchestrates $\pi_{\mathcal{A}}$ and executes $\epsilon_{\mathcal{A}}$ pipeline $P$ but also autonomously discovers task $T'$ to begin with. The human $\mathcal{H}$ transitions to a passive recipient of the output:
\vspace{-0em}
\begin{equation*}
\begin{aligned}
\mathcal{A} &: \text{Discover}_{\mathcal{A}}(\mathcal{D}, \mathcal{E}, \mathcal{M}) \rightarrow \mathcal{T}'; \\[-1pt]
&\quad \pi_{\mathcal{A}}(\mathcal{T}', \mathcal{D}, \mathcal{E}, \mathcal{M}) \rightarrow P; ~~\epsilon_{\mathcal{A}}(P, \mathcal{D}, \mathcal{E}, \mathcal{M}) \rightarrow \mathcal{O} \\[-1pt]
\mathcal{H} &: \text{Receive}(\mathcal{O})
\end{aligned}
\end{equation*}
\vspace{-0em}

This leap is associated with three core advancements: the capacity for autonomous problem discovery, the establishment of trustworthy self-governance, and long-horizon views. 

\ptitle{Autonomous Problem Discovery.} 
To achieve autonomous problem discovery, L4 data agents are expected to demonstrate critical reasoning, task-oriented awareness, and curiosity to investigate data throughout its lifecycle.
Beyond execution, they must critically evaluate data to identify gaps, anomalies, and latent issues worthy of investigation, such as detecting a recurring drop in transactions while cleaning sales data. 
This transforms the data pipeline from a passive flow into an active source of analytical hypotheses.

\ptitle{Trustworthy Self-governance}. 
To effectively address the diverse and unforeseen challenges, L4 data agents must operate as reliable generalists, fully entrusted with data-related responsibilities beyond L3's supervised scope, which demands a comprehensive mastery of data lifecycle. Their ability to self-manage resources and ensure operational accuracy and security is foundational to trust, making human oversight unnecessary.

\ptitle{Long-Horizon and Holistic Views.} 
High autonomy requires L4 data agents to possess long-term planning and decision-making capabilities. They must optimize across processes and make strategic trade-offs, such as balancing immediate data cleaning costs against long-term analytical benefits. This foresight prioritizes holistic benefits and long-term objectives over short-term gains, enabling the management of goals and resources over extended periods. Thus, L4 data agents are expected to adopt a long-horizon and holistic perspective that goes beyond tactical operations and local optimizations.

\begin{figure}[t!]
    \centering
    \includegraphics[width=\linewidth]{\paperroot 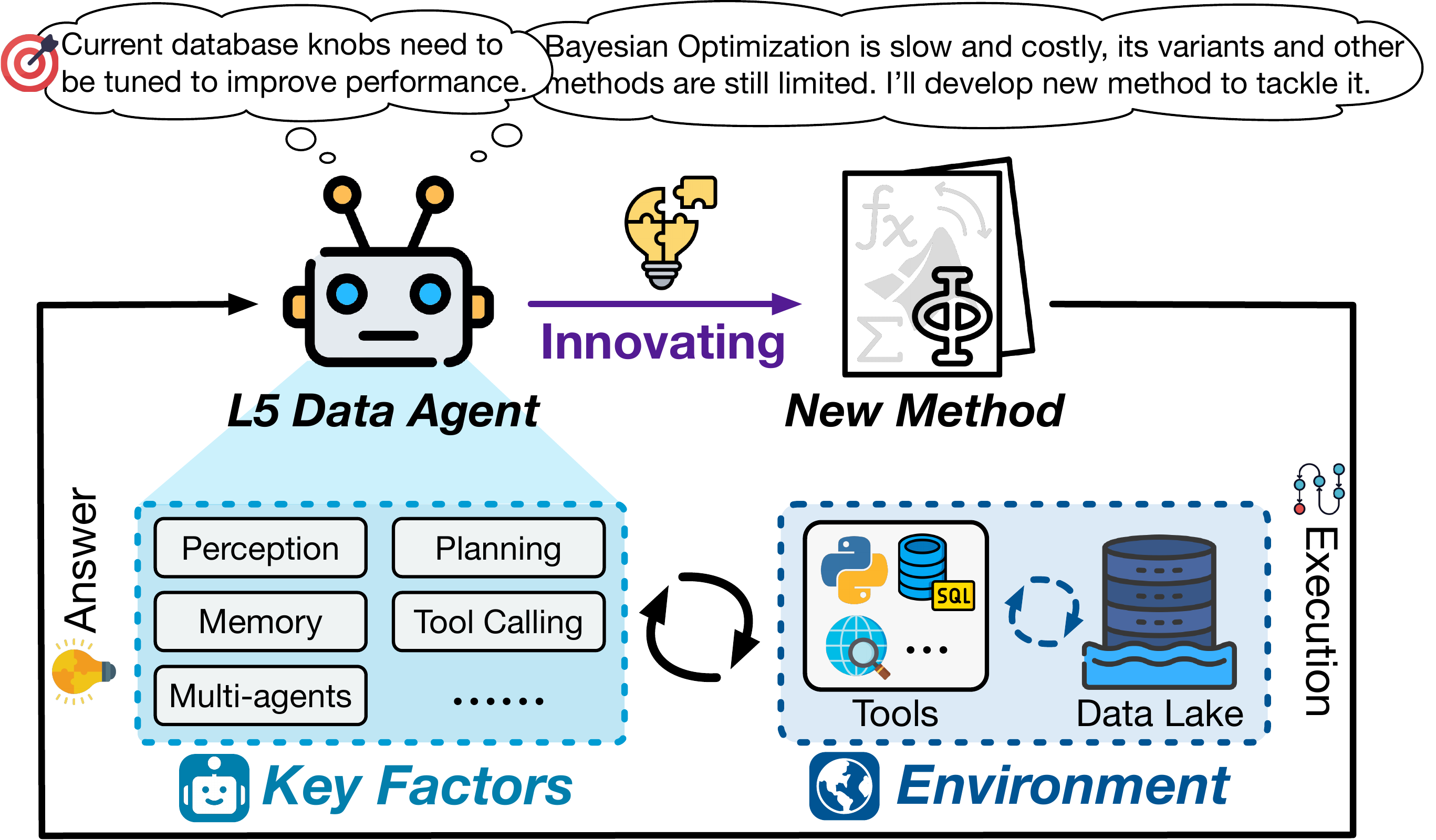}
    \vspace{-0em}
    \caption{L5 Data Agents (Full Autonomy).}
    \vspace{-0em}
    \label{fig:l5_agent}
\end{figure}

\vspace{-0em}
\subsection{L5: The Ultimate Vision of Ubiquitous and Generative Data Agents}

L5 represents the ultimate vision: a fully autonomous data agent functioning as a ubiquitous and generative data scientist. Beyond merely applying existing methods, it actively creates new knowledge by identifying when conventional approaches are insufficient and pioneering novel solutions, as illustrated in Figure~\ref{fig:l5_agent}. For example, an L5 agent might develop new sampling theories, design federated data preparation frameworks, or invent original visualization grammars.

Formally, the data agent $\mathcal{A}$ not only autonomously identifies the promising task $\mathcal{T}'$ but also invents a new, innovative process $\Phi$ (\eg a new theory, algorithm, or paradigm) rather than relying on existing methods, while human $\mathcal{H}$ disengages:
\vspace{-0em}
\begin{equation*}
\begin{aligned}
\mathcal{A} &: \text{Discover}_{\mathcal{A}}(\mathcal{D}, \mathcal{E}, \mathcal{M}) \rightarrow \mathcal{T}'; \\[-1pt]
& \quad \text{Innovate}_{\mathcal{A}}(\mathcal{T}', \mathcal{D}, \mathcal{E}, \mathcal{M}) \rightarrow \Phi; \\[-1pt]
& \quad \Phi(\mathcal{T}', \mathcal{D}, \mathcal{E}, \mathcal{M}) \rightarrow \mathcal{O} \\[-1pt]
\mathcal{H} &: \emptyset
\end{aligned}
\end{equation*}
\vspace{-0em}

This final stage transcends technical proficiency to embrace innovation, where data agents establish state-of-the-art algorithms and models, pioneer new analytical paradigms, even formulate groundbreaking theories. At this visionary stage, human engagement becomes unnecessary and potentially detrimental, as L5 data agents are envisioned as fully autonomous and generative intellects with unparalleled expertise, pushing the frontiers of data management, preparation, and analysis.

\vspace{-0em}
\subsection{The Research Odyssey Ahead}

The progression from Proto-L3 data agents to L3, L4, and L5 is an odyssey, requiring fundamental breakthroughs rather than incremental improvements. 
As discussed, key challenges include developing agents with autonomous orchestration, versatility, critical reasoning, intrinsic motivation for independent task discovery, long-horizon planning, and robust safety guarantees.
While realizing L4 and L5 remains a long-term vision, we believe exploratory efforts, such as developing data agents that can autonomously manage diverse tasks within data lakes over extended periods with progressively reduced human intervention, constitute essential and achievable steps toward the eventual goal of fully autonomous data agents.

\vspace{-0em}
\section{Conclusion}
\label{sec:conclusion}
\vspace{0em}

In this survey, we introduce the first systematic hierarchical taxonomy for data agents, comprising six levels that delineate progressive shifts in autonomy from manual operations to fully autonomous systems. We conduct a structured literature review through this taxonomy, 
and further analyze evolutionary leaps and technical gaps between levels, with emphasis on the ongoing L2-to-L3 transition. Finally, we provide a roadmap for future research, outlining opportunities to bridge gaps and envision higher autonomy toward proactive and generative data agents (L4--L5).

\vspace{-.0em}

\bibliographystyle{IEEEtran}
\bibliography{reference}

\end{document}